\documentclass[pdflatex,sn-mathphys-ay]{sn-jnl}% Math and Physical Sciences Author Year Reference Style % original
\usepackage{graphicx}%
\usepackage{multirow}%
\usepackage{amsmath,amssymb,amsfonts}%
\usepackage{amsthm}%
\usepackage{mathrsfs}%
\usepackage[title]{appendix}%
\usepackage{xcolor}%
\usepackage{textcomp}%
\usepackage{manyfoot}%
\usepackage{booktabs}%
\usepackage{algorithm}%
\usepackage{algorithmicx}%
\usepackage{algpseudocode}%
\usepackage{listings}%
\theoremstyle{thmstyleone}%
\theoremstyle{thmstyletwo}%

\theoremstyle{thmstylethree}%

\begin{document}

\title[Search for submicrosecond optical transients with the TAIGA-HiSCORE Cherenkov array]{Wide-field search for submicrosecond optical and gamma transients as manifestation of primordial black holes evaporation with the TAIGA-HiSCORE Cherenkov array}


% \author{\fnm{A.~D.}~\sur{Panov}}
% \email{panov@dec1.sinp.msu.ru}
% \affil{%
% Lomonosov Moscow State University Skobeltsyn Institute of Nuclear Physics (MSU SINP),
% Leninskie gory 1(2), GSP-1, Moscow, 119991, Russia.
% }


\author[2]{\fnm{I.~I.}~\sur{Astapov}}
\author[3]{\fnm{G.~M.}~\sur{Beskin}}
\author[4]{\fnm{P.~A.}~\sur{Bezyazykov}}
\author[5]{\fnm{A.~V.}~\sur{Blinov}}
\author[1]{\fnm{E.~A.}~\sur{Bonvech}}
\author[5]{\fnm{A.~N.}~\sur{Borodin}}
\author[4]{\fnm{N.~M.}~\sur{Budnev}}
\author[1]{\fnm{A.~V.}~\sur{Bulan}}
\author[4]{\fnm{P.}~\sur{Busygina}}
\author[1]{\fnm{D.~V.}~\sur{Chernov}}
\author[6]{\fnm{A.}~\sur{Chiavassa}}
\author[4]{\fnm{A.~N.}~\sur{Dyachok}}
\author[4]{\fnm{A.~R.}~\sur{Gafarov}}
\author[7,8]{\fnm{A.~Yu.}~\sur{Garmash}}
\author[4]{\fnm{E.~O.}~\sur{Gress}}
\author[4]{\fnm{O.~A.}~\sur{Gress}}
\author[4]{\fnm{T.~I.}~\sur{Gress}}
\author[5]{\fnm{A.~A.}~\sur{Grinyuk}}
\author[4]{\fnm{O.~G.}~\sur{Grishin}}
\author[4]{\fnm{A.~L.}~\sur{Ivanova}}
\author[4]{\fnm{A.~D.}~\sur{Ivanova}}
\author[4]{\fnm{M.~A.}~\sur{Ilyushin}}
\author[1]{\fnm{N.~N.}~\sur{Kalmykov}}
\author[5]{\fnm{Kh.}~\sur{Karatas}}
\author[4]{\fnm{V.~V.}~\sur{Kindin}}
\author[4]{\fnm{S.~N.}~\sur{Kiryuhin}}
\author[4]{\fnm{K.~G.}~\sur{Kompaniets}}
\author[1]{\fnm{E.~E.}~\sur{Korosteleva}}
\author[1]{\fnm{V.~A.}~\sur{Kozhin}}
\author[7,8]{\fnm{E.~A.}~\sur{Kravchenko}}
\author*[1]{\fnm{L.~A.}~\sur{Kuzmichev}}\email{kuz@dec1.sinp.msu.ru}
\author[1]{\fnm{A.~P.}~\sur{Kryukov}}
\author[9]{\fnm{A.~A.}~\sur{Lagutin}}
\author[4]{\fnm{Yu.}~\sur{Lemeshev}}
\author[10]{\fnm{B.~K.}~\sur{Lubsandorzhiev}}
\author[1]{\fnm{N.~B.}~\sur{Lubsandorzhiev}}
\author[10]{\fnm{S.~B.}~\sur{Lubsandorzhiev}}
\author[10]{\fnm{A.~D.}~\sur{Lukanov}}
\author[4]{\fnm{S.~D.}~\sur{Malakhov}}
\author[4]{\fnm{R.~R.}~\sur{Mirgazov}}
\author[4]{\fnm{R.~D.}~\sur{Monkhoev}}
\author[1]{\fnm{E.~A.}~\sur{Osipova}}
\author[4]{\fnm{A.~L.}~\sur{Pakhorukov}}
\author[1]{\fnm{A.~T.}~\sur{Panferova}}
\author[4]{\fnm{L.~V.}~\sur{Pankov}}
\author*[12]{\fnm{A.~D.}~\sur{Panov}}\email{panovenator@gmail.com}
\author[2]{\fnm{A.~A.}~\sur{Petrukhin}}
\author[1]{\fnm{D.~A.}~\sur{Podgrudkov}}
\author[1]{\fnm{E.~G.}~\sur{Popova}}
\author[1]{\fnm{E.~B.}~\sur{Postnikov}}
\author[1]{\fnm{V.~V.}~\sur{Prosin}}
\author[4]{\fnm{A.~A.}~\sur{Pushnin}}
\author[9]{\fnm{R.~I.}~\sur{Raikin}}
\author[1]{\fnm{A.~Yu.}~\sur{Razumov}}
\author[4]{\fnm{E.}~\sur{Rjabov}}
\author[10]{\fnm{G.~I.}~\sur{Rubtsov}}
\author[4]{\fnm{V.~S.}~\sur{Samoliga}}
\author[5]{\fnm{A.~V.}~\sur{Shaikovsky}}
\author[10]{\fnm{A.~Yu.}~\sur{Sidorenkov}}
\author[1]{\fnm{A.~A.}~\sur{Silaev}}
\author[1]{\fnm{A. A.}~\sur{Silaev (junior)}}
\author[1]{\fnm{A.~V.}~\sur{Skurikhin}}
\author[7,8]{\fnm{A.~V.}~\sur{Sokolov}}
\author[1]{\fnm{L.~G.}~\sur{Sveshnikova}}
\author[4]{\fnm{V.~A.}~\sur{Tabolenko}}
\author[4]{\fnm{A.~B.}~\sur{Tanaev}}
\author[4]{\fnm{M.}~\sur{Ternovoy}}
\author[1]{\fnm{P.~A.}~\sur{Volchugov}}
\author[9]{\fnm{N.~V.}~\sur{Volkov}}
\author[10]{\fnm{D.~M.}~\sur{Voronin}}
\author[2]{\fnm{I.~I.}~\sur{Yashin}}
\author[4]{\fnm{A.~V.}~\sur{Zagorodnikov}}
\author[4]{\fnm{D.~P.}~\sur{Zhurov}}
\author[11]{\fnm{V.~N.}~\sur{Zirakashvili}}

%*************************************************************************************************************

\affil[1]{%
\orgname{Lomonosov Moscow State University Skobeltsyn Institute of Nuclear Physics (MSU SINP)},
\orgaddress{\street{Leninskie gory 1(2), GSP-1}, \city{Moscow}, \postcode{119991}, \country{Russia}}
}
\affil[2]{%
\orgname{National Research Nuclear University MEPhI (Moscow Engineering Physics Institute)},
\orgaddress{\street{Kashirskoe highway 31}, \city{Moscow}, \postcode{115409}, \country{Russia}}
}
\affil[3]{%
\orgname{Special Astrophysical Observatory}, \orgaddress{\city{Nizhnij Arkhyz}, \state{Zelenchukskiy region, Karachai-Cherkessian Republic}, \postcode{369167}, \country{Russia}}
}
\affil[4]{%
\orgdiv{Institute of Applied Physics}, \orgname{Irkutsk State University (API ISU)}, \orgaddress{\street{Gagarin Blvd. 20}, \city{Irkutsk},
\postcode{664003}, \country{Russia}}
}
\affil[5]{%
\orgname{Joint Institute for Nuclear Research}, \orgaddress{\street{Joliot-Curie 6}, \city{Dubna}, \state{Moscow Region}, \postcode{141980}, \country{Russia}}
}
\affil[6]{%
\orgname{Physics Department of the University of Torino and the National Institute of Nuclear
Physics INFN}, \orgaddress{\postcode{10125}, \city{Torino}, \country{Italy}}
}%
\affil[7]{%
\orgname{Novosibirsk State University}, \orgaddress{\street{Pirogova 1}, \city{Novosibirsk}, \postcode{630090}, \country{Russia}}
}
\affil[8]{%
\orgname{Budker Institute of Nuclear Physics of the Siberian Branch of the Russian Academy of
Sciences}, \orgaddress{\street{Lavrentyev Prosp. 11}, \city{Novosibirsk}, \postcode{630090}, \country{Russia}}
}
\affil[9]{%
\orgname{Altai State University}, \orgaddress{\street{Lenina 61}, \city{Barnaul}, \postcode{656049}, \country{Russia}}
}
\affil[10]{%
\orgname{Institute for Nuclear Research of the Russian Academy of Sciences}, \orgaddress{\street{60th October
Anniversary 7a}, \postcode{117312}, \city{Moscow}, \country{Russia}}
}
\affil[11]{%
\orgname{Pushkov Institute of Terrestrial Magnetism, Ionosphere and Radio Wave Propagation of the
Siberian Branch of the Russian Academy of Sciences (IZMIRAN)}, \orgaddress{\street{Kaluzhskoe highway 4},
\city{Moscow, Troitsk}, \postcode{108840}, \country{Russia}}
}%
\affil[12]{\orgname{Free researcher}, \orgaddress{\city{Moscow}, \country{Russia}}}


\abstract{We demonstrated in our papers \cite{PANOV-2021,PANOV-2025}, that the TAIGA-HiSCORE Cherenkov array physics, originally designed for research in cosmic rays and ultra-high-energy gamma-ray astronomy, can also be used for wide-field searches for very rare submicrosecond optical transients of astrophysical origin. The papers \cite{PANOV-2021,PANOV-2025} presented the results of the first four seasons of optical transient searches (2018--2022) and the resulting upper limits on the flux of corresponding events. The present paper presents data processing from two new seasons of optical transient searches (2022--2024). This paper further develops a method for searching for narrow spatiotemporal groups of ultra-high-energy gamma-rays (above 100\,TeV). Since the evaporation of relic black holes can be accompanied by both by the emission of a short light pulse and by a flash of ultra-high-energy gamma-rays, a multimessenger technique for searching for black hole evaporation was in fact implemented in this work, where two different methods are implemented within a single instrument. The search for narrow spatiotemporal groups of gamma-rays was carried out in time windows of 1000\,$\mu$s, 100\,$\mu$s, and 6\,$\mu$s. Since narrow groups of gamma were not detected, the upper limit on their frequency were obtained as $10^{-3}\,\mathrm{ster}^{-1}\mathrm{h}^{-1}$. The upper limit on the number of submicrosecond optical transients have also been updated: $10^{-4}\mathrm{erg/s/cm}^2$ as $5\cdot10^{-4}\,\mathrm{ster}^{-1}\mathrm{h}^{-1}$.}

\keywords{Cherenkov array, submicrosecond optical tranzients, gamma-ray astronomy, primordial black holes, Hawking radiation}



\maketitle

\section{Introduction}
\label{INTRODUCTION}

The TAIGA-HiSCORE Cherenkov array is one of the instruments of the TAIGA astrophysical complex, designed for research in the field of gamma-ray astronomy and ultrahigh-energy cosmic rays physics using extensive air showers (EAS) \cite{TAIGA-NIM-2017A,TAIGA-NIM-2017B,TAIGA-NIM-2020A,TAIGA-NIM-2020B}. The complex is located in the Tunka Valley, 50 km from the Lake Baikal. Several of our previous studies have demonstrated that the TAIGA-HiSCORE array is also suitable for searching for submicrosecond optical transients (10~ns--150~ns) of astrophysical origin \cite{PANOV-2021,PANOV-2025}, i.e., it can be used as a tool for optical astronomy. An essential feature of the TAIGA-HiSCORE array compared to conventional optical telescopes is the ability to search for very rare events due to the instrument's wide field of view, which is on the scale of one steradian.

Due to the astroclimatic conditions in the Tunka Valley the TAIGA astrophysical complex carryes out measurements from approximately mid-autumn, then in winter, and up to mid-spring each year. These periods are referred to as the TAIGA winter seasons. The search for astrophysical optical transients using the TAIGA-HiSCORE array has been ongoing since the 2018--2019 winter season. The results of the 2018--2019 season are published in the paper \cite{PANOV-2021}, and the results of the three seasons 2019--2020, 2020--2021, and 2021--2022 are published in the paper \cite{PANOV-2025}. No candidates to submicrosecond optical transient were found up to now. This new paper presents the results of optical transient searches in the new HiSCORE  seasons 2022--2023 and 2023--2024. Additionally, the method was extended to include a search for gamma-ray bursts that can accompany the optical flashes of the evaporation of primordial black holes.

The physical motivation for searching for astrophysical optical transients in the submicrosecond range of durations is discussed in detail in our article \cite{PANOV-2025}, where one can also find references to the relevant literature. The three main sources of such optical transients considered in \cite{PANOV-2025} were: 1) flashes of light associated with the evaporation of the remnants of primordial black holes; 2) stochastic variability of the radiation of the plasma accreting onto the stellar mass black holes; 3) laser pulses from an extraterrestrial civilization.

Some notes regarding primordial black hole evaporation should be done here. The final stages of a black hole evaporation cannot currently be fully understood, as they must be described within the framework of a theory of quantum gravity, that is not yet known. We emphasize that, despite the insufficient understanding of the processes accompanying the final stages of black hole evaporation, there is reason to believe that the final explosion, completing the evaporation, will be accompanied, among other signatures, by an optical flash. Such a flash may be caused, in particular, by synchrotron radiation from high-energy charged particles emitted by the black hole remnant into the interstellar magnetic field, where the primordial black hole was immersed before the explosion \cite{REES-1977,BLANDFORD-1977,PETKOV-2019}.

Another signature of a black hole evaporation could be groups of high-energy gamma rays (multi-TeV energies) that arrive from a single point in the sky within a short time interval. High-energy gamma ray astronomy is the primary purpose of the TAIGA-HiSCORE array. Since the array detects not only the occurrence of a gamma ray but also its direction of arrival, such narrow spatiotemporal groups of gamma rays can be easily identified by TAIGA-HiSCORE if they occur.

The search for the evaporation of black holes using such groups of gamma rays has already been carried out in a number of works \cite{CYGNUS-1993,Tibet-1995,Whipple-2006,PETKOV-2008,PETKOV-2009A,PETKOV-2009B,Milagro-2015,VERITAS-2017,FERMI-2018,HAWC-2020,HESS-2023}. However, as noted in \cite{PETKOV-2019}, detecting black hole evaporation simultaneously using several different methods would be much more interesting and useful. It was proposed in \cite{PETKOV-2019} to use several \emph{different} instruments simultaneously for this purpose, but the TAIGA-HiSCORE array allows for a multimessenger approach to be implemented within a \emph{single} instrument. Using various event signatures, HiSCORE allows for the detection and discrimination of both narrow spatiotemporal groups of gamma-ray photons and optical flashs that can accompany black hole evaporation. This is a multimessenger method inplemented within a single instrument.

The aim of this work is not only to search for submicrosecond optical transients of astrophysical origin, regardless of their nature, but also to implement a multimesenger approach to searching for the evaporation of primordial black holes within the framework of a single TAIGA-HiSCORE instrument. The article consists of two relatively independent parts. One examines the search for submicrosecond optical transients (Section~\ref{OPT}), and the other examines the search for narrow spatiotemporal groups of EAS events (Section~\ref{GG}), which can be interpreted as groups of gamma-ray quanta accompanying the evaporation of black holes. Each Section \ref{OPT} and \ref{GG} of the paper describes the corresponding methodology and results. Section~\ref{DATA} presents a description of the HiSCORE instrument and the 2022--2023 and 2023--2024 season data common for the entire paper, in Section~\ref{CONCLUSION} the main results are discussed.

\section{The HiSCORE instrument and the data of 2022--2023 and 2023--2024 seasons}
\label{DATA}

The HiSCORE (High Sensitivity Cosmic Rays and Gamma Explorer) array consists of 120 Cherenkov optical stations, each with four photomultiplier tubes (PMTs) with a total entrance pupil area of approximately 0.5\,m$^2$. All four PMTs of the optical station have common field of view. It is difficult to determine the exact size of the field of view of optical station because it depends on the brightness of the flash of light, but it is a value of the sclale of one steradian. The stations form a grid with a step of 106 m, which covers an area slightly greater than one square kilometer. All optical stations are oriented identically, but their orientation may vary from season to season. The integration time of the optical station electronics is 10~ns, and the maximum duration of the recorded optical signal is set at 150~ns (this value is adjustable). Optical stations are sensitive to the visible range of light. The station detection threshold is approximately 3000 green light quanta per square meter within an integration time of 10~ns, which roughly corresponds to an object with a brightness of -4m (Venus at its highest brightness). The design and construction details of the HiSCORE array are presented in the papers \cite{HISCORE-2011, TAIGA-NIM-2017B}. In the main operating mode of the HiSCORE instrument, the optical stations measure the amplitudes of the light signal caused by the Cherenkov light of the EAS, as well as the temporal evolution of the Cherenkov light front, reconstructed from the stations time triggers sequence. It is possible from this data to reconstruct the arrival direction of a cosmic particle (charged particle or gamma ray), the position of the shower axis, its initial energy, and several other important EAS characteristics.

\begin{figure}
\centering
\includegraphics[width=0.49\textwidth]{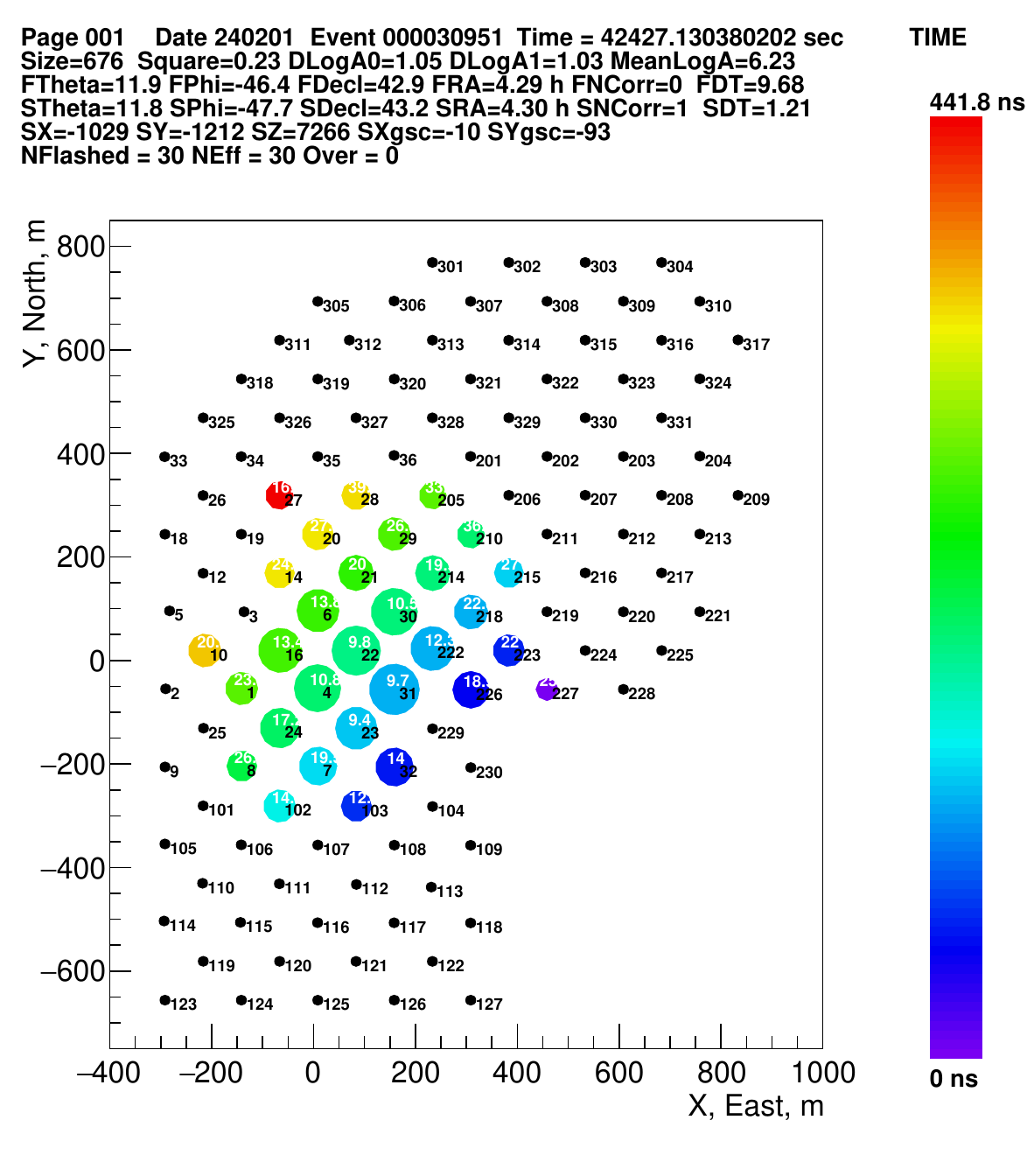}
\includegraphics[width=0.49\textwidth]{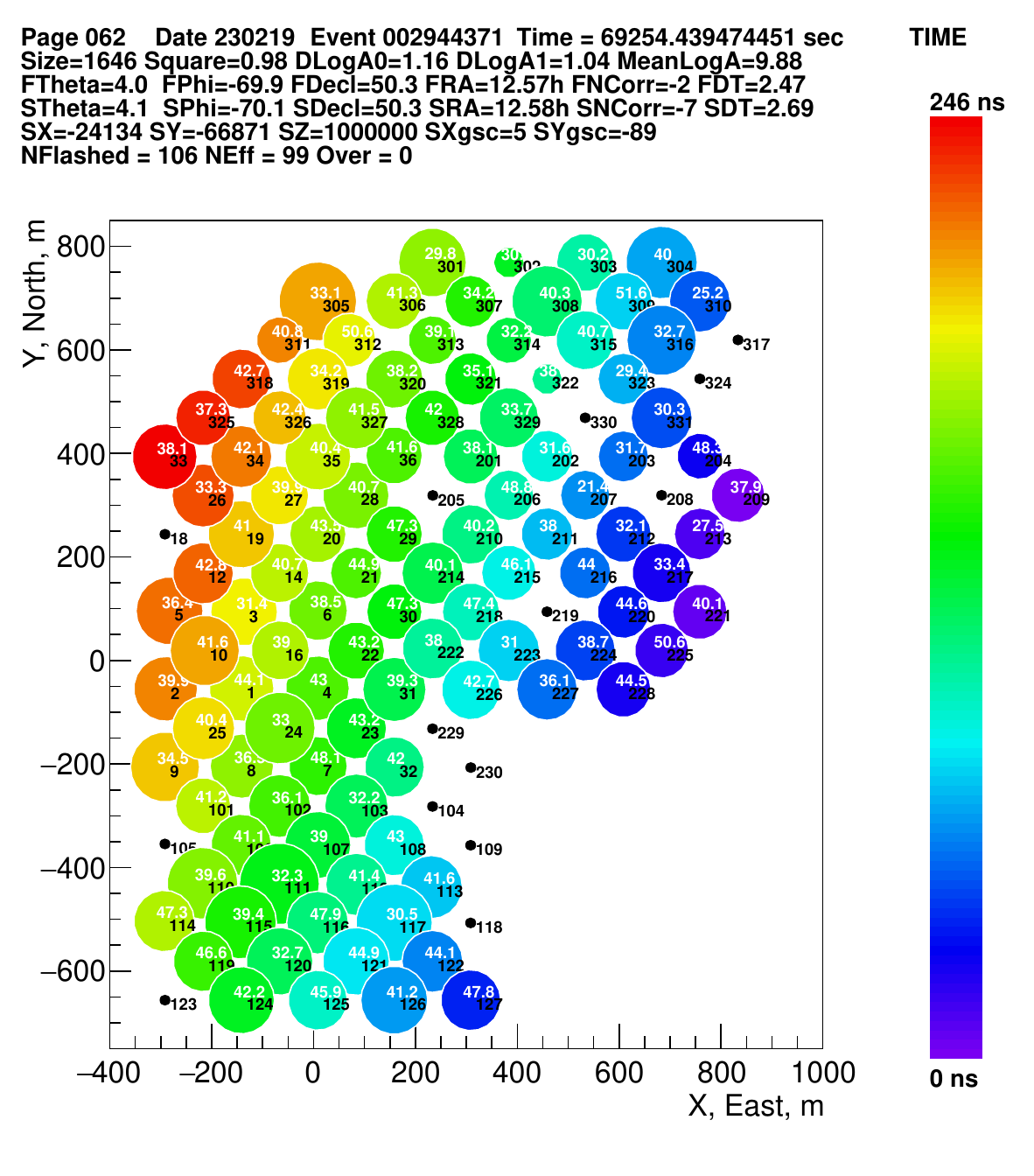}
\caption{A ``portrait'' of generic EAS event detected by the HiSCORE array and event from the lidar of Daqi-1 sattelite. See text for explanation}\label{FigGenericEAS}
\end{figure}

Fig.~\ref{FigGenericEAS} (left panel) shows a generic EAS event from the 2023--2024 season, detected by the HiSCORE array. The figure shows the locations of HiSCORE optical stations, with colored circles indicating the optical stations that have been triggered in the event. The circle square is proportional to the logarithm of the signal amplitude, the white numbers inside the circles indicate the optical station's signal duration in nanoseconds, and the circle color encodes the optical station's trigger time. The black numbers represent the optical station numbers adopted by the TAIGA collaboration. The top portion of the ``event portrait'' contains a list of the event parameters obtained during the data processing. Among the parameters there is the direction of the shower axis, reconstructed by fitting the light front of EAS with a flat front (FTheta=11.9 FPhi=-46.4 FDecl=42.9 FRA=4.29~h in Fig.~\ref{FigGenericEAS}, left panel) and with a spherical front (STheta=11.8 SPhi=-47.7 SDecl=43.2 SRA=4.30~h), in both cases in local and equatorial coordinates. Theta means zenith angle, Phi means snadard geographical azimuth, both in degrees; Decl means declination (degrees) and RA means right accession (hours). Prefix F relates to a flat front reconstruction, prefix S relates to a spherical front reconstruction. There are some other parameters, which we will not discuss here. It can be seen that the EAS events generally have the appearance of a ``spot'' located somewhere within the area occupied by the array. The shower axis passes near the center of the spot and it is clearly seen that the amplitudes of the signals of optical stations near the shower axis are greater than at the periphery of the shower.

The main calibration procedures of the HiSCORE instrument include amplitude calibration of optical station signals and correction of optical station trigger timing errors. The goal of amplitude calibration is to equalize the amplitude distributions of optical station signals. During each night, all optical stations should produce identical signal amplitude distributions, indicating identical gains of the optical stations. The identity of the amplidute distributions of different optical stations permits to compare correctly the strengths of the signals of different stations within each event. Correction of optical station trigger timing errors eliminates small uncertainties in the time delays of optical communication links and is performed using LED signals from drones passage the HiSCORE array. The calibration methods used in this work were identical to those described in detail in the paper \cite{PANOV-2025}, so we will not dwell on the details of these methods here.

Unlike the EAS arrays, which detect charged particles and can operate 24/7 year-round, the HiSCORE array, which relies on EAS Cherenkov light detection, can only operate on clear, moonless nights. The total HISCORE exposure during the 2022--2023 winter season was 696.7~hours, with a total of 248,752,575 events detected. During the 2023--2024 winter season, the exposure was 659.3~hours, with a total of 268,145,330 events detected.

The time signal sequence of each optical station is recorded independently, and then the events records are generated off-line. The sign of an event was the fact that at least two optical stations were triggered within a 6.5\,$\mu$s time window. The size of the time window is determined by the linear dimensions of the HiSCORE array. Thus, the minimum ``multiplicity'' of an event is two optical stations. The statistics presented above apply to the events with a multiplicity of at least two. However, for an event with a multiplicity of two, it is impossible to reconstruct the shower's axis direction. The minimum multiplicity of an event for which the reconstruction of direction is possible is three. In this case, reconstruction can be performed using a plane front approximation, but it is impossible to estimate the accuracy of the approximation, since the plane is drawn precisely and uniquely by three points. The inability to determine the accuracy of the approximation creates troubles for the problems considered in this paper. Actually events with a multiplicity of at least five were used for the final analysis. For such events, either a flat-front approximation or a spherical-front approximation can be used to reconstruct the event axis direction. In both cases, the quality of the approximation can be assessed by the deviation of the actual station trigger times from the fitting surface. For the winter 2022--2023 season, 81012017 events with a multiplicity of at least five stations were detected, and for the 2023--2024 season, 87994903 such events were detected. The number of events with a multiplicity of at least five is approximately one third of the total number of detected events with a multiplicity of at least two.

\section{Search for submicrosecond optical transients}
\label{OPT}

Two main signatures were used to select potential candidates for distant optical transients in this paper: the event area signature and the signature of the altitude $Z$ of the center of the event's reconstructed spherical light front. These are exactly the same signatures that were used in our previous paper \cite{PANOV-2025}.

The event area is the area of the minimal convex polygon in the $(X,Y)$ plane within which the event can be inscribed, with the area expressed in units relative to the area of the entire HiSCORE array, defined in the same way.

The idea behind using the event area signature is that a general EAS event has the shape of a ``spot'' in the plane of the HiSCORE array (see Fig.\,\ref{FigGenericEAS}, left panel), while a distant optical transient event should cover the entire array. The right panel of Fig.\,\ref{FigGenericEAS} shows one of the events generated by the lidar on board the Daqi-1 satellite\footnote{Krebs, Gunter D. "Daqi 1 (DQ 1, AEMS 1)", Gunter's Space Page.,\\ \texttt{https://space.skyrocket.de/doc\_sdat/daqi-1.htm}}. Since the satellite flies at an altitude of about 700\,km, the laser light front at the HiSCORE level is very close to flat one and therefore well models the light front of a distant astrophysical optical transient. Clearly, an EAS event, as a rule, will have a significantly smaller area than events of a distant optical transient, which allows the latter to be distinguished from the EAS background.

\begin{figure}
\centering
\includegraphics[width=0.6\textwidth]{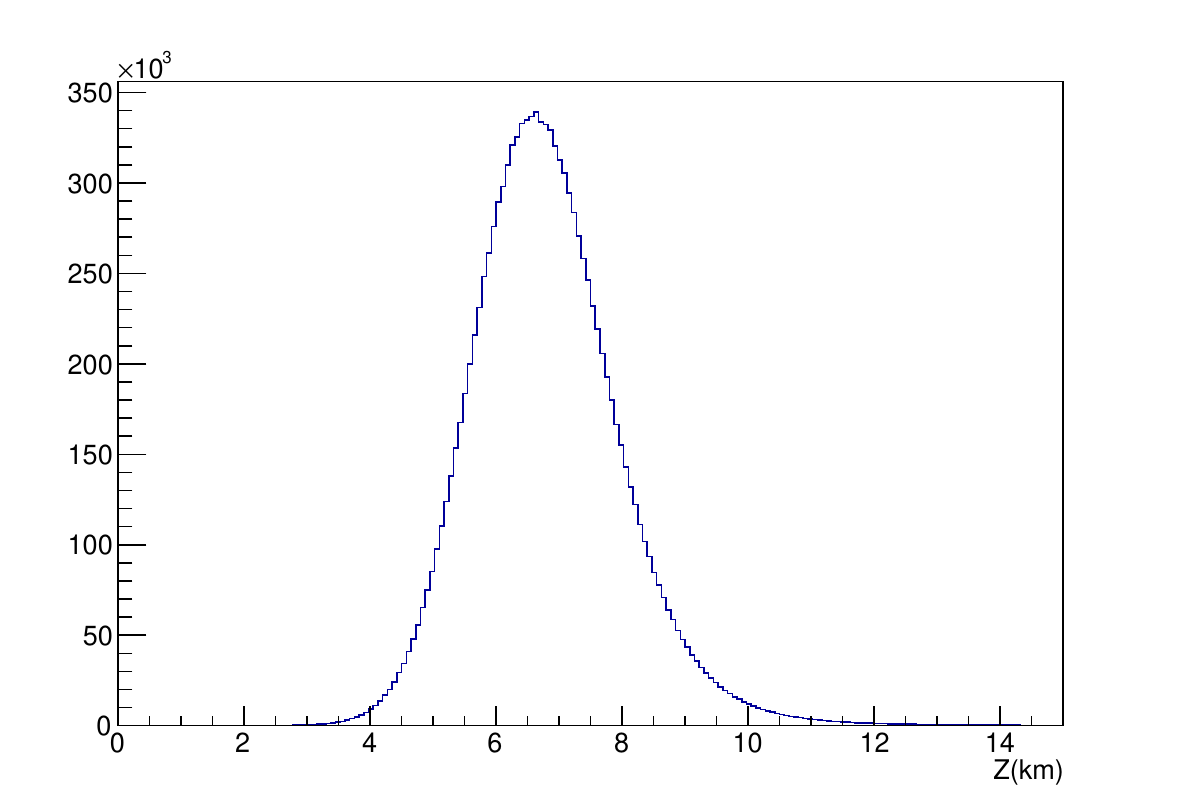}
\caption{Distribution of the reconstructed altitude $Z$  of the spherical front center above the HiSCORE array plane for EAS events with a multiplicity of at least 12 triggered optical stations, for the data of 2023-2024 season.}\label{FigZ-2024}
\end{figure}

Reconstruction of the shower's temporal structure by a spherical light front yields three coordinates of the light front center  in the HiSCORE array coordinate system $X,Y,Z$. Here the $Z$ coordinate represents the latitude of the front center above the plane tangent to the Earth's surface at the array coordinate origin. For the great majority of EAS events, the $Z$ value is less than 15\,km, see Fig.~\ref{FigZ-2024}. The Daqi-1 satellite events have nearly flat light fronts, so the satellite's flight altitude is reconstructed from the light front curvature with a large error, but this value always turns out to be greater than 30 km. Sometimes the algorithm cannot find $Z$ for a sattelite event within the upper limit 1000\,km of the search area, and then the algorithm reports $Z = 1000$\,km. Since satellite events are very similar to events of a true astrophysical optical transient, then $Z > 30$\,km is a very good signature for such events.

\begin{figure}
\centering
\includegraphics[width=0.49\textwidth]{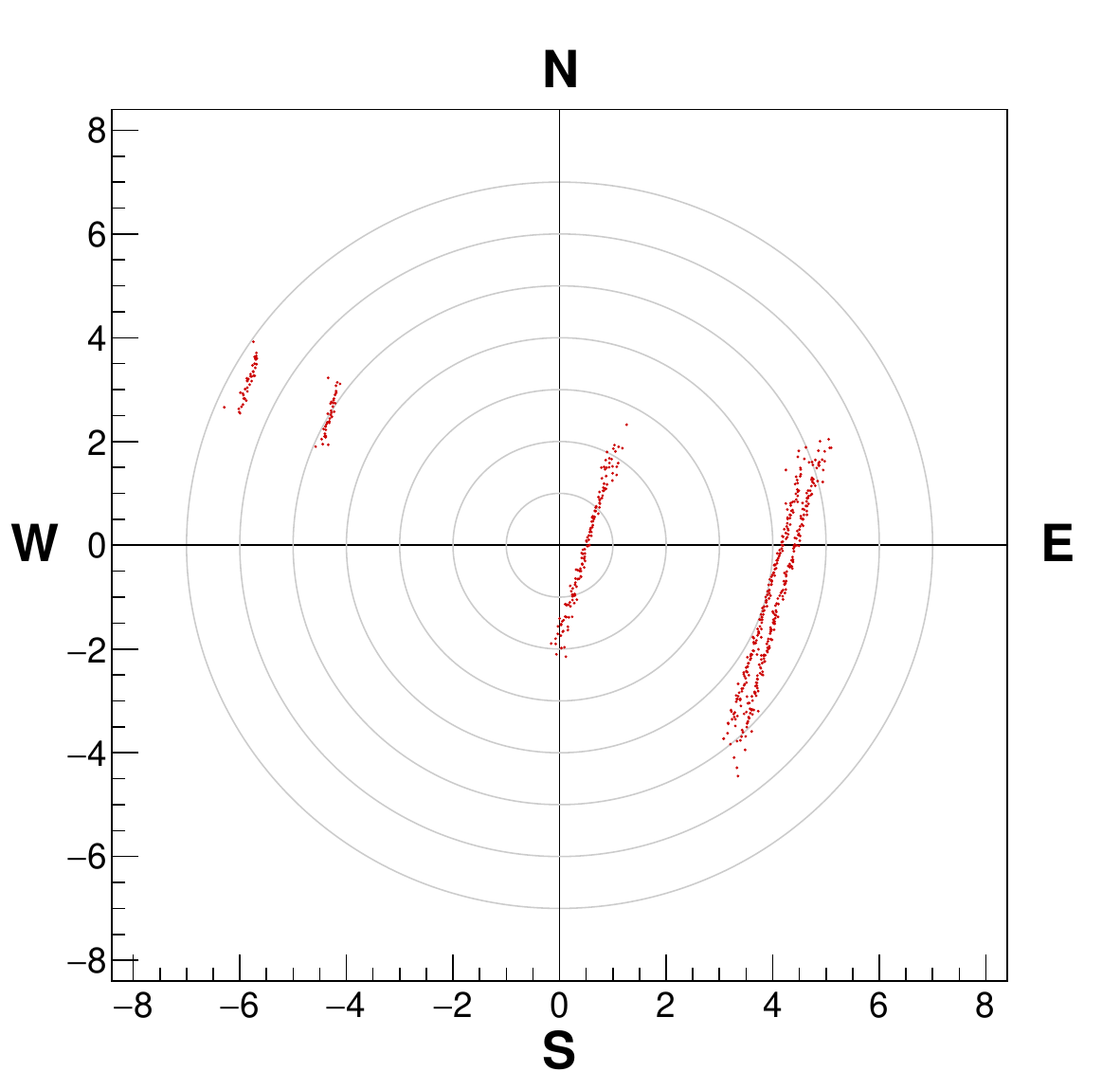}
\includegraphics[width=0.49\textwidth]{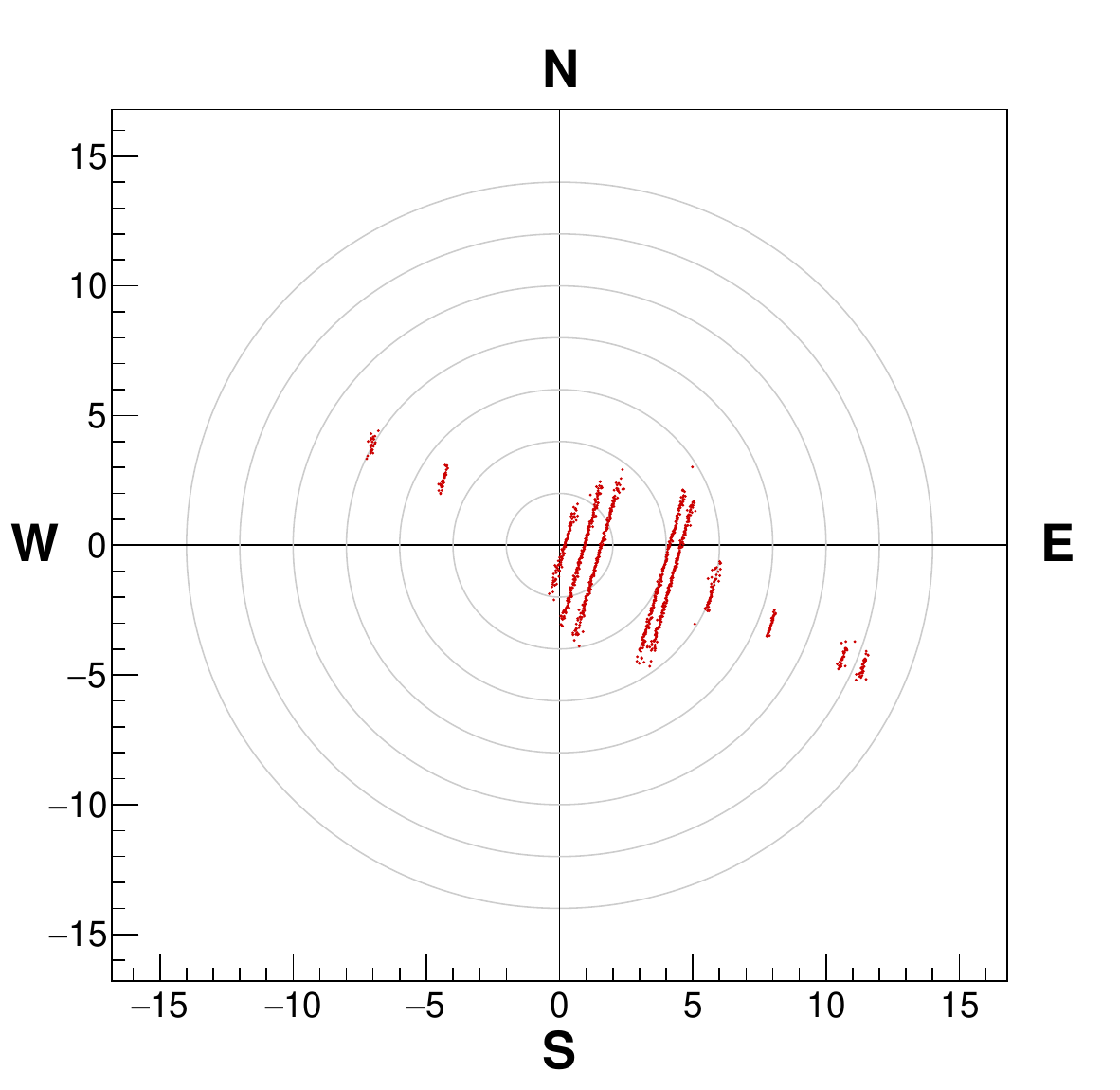}
\caption{Daqi-1 satellite passage trajectories over the HiSCORE array in the 2022--2023 and 2023--2024 seasons. The trajectories are shown in the orthogonal angular coordinate system with the origin at the zenith. Angles are shown in degrees.}\label{FigSatt}
\end{figure}

To fine-tune the filters for selecting candidates for optical transients of astrophysical origin, Daqi-1 events were used as a prototype. Several satellite passage the HiSCORE array were detected in each of the 2022--2023 and 2023--2024 seasons (see Fig.~\ref{FigSatt}). The satellite's events provide sufficient statistics for filters tuning. The filter tune method was as follows.

\begin{figure}
\centering
\includegraphics[width=0.49\textwidth]{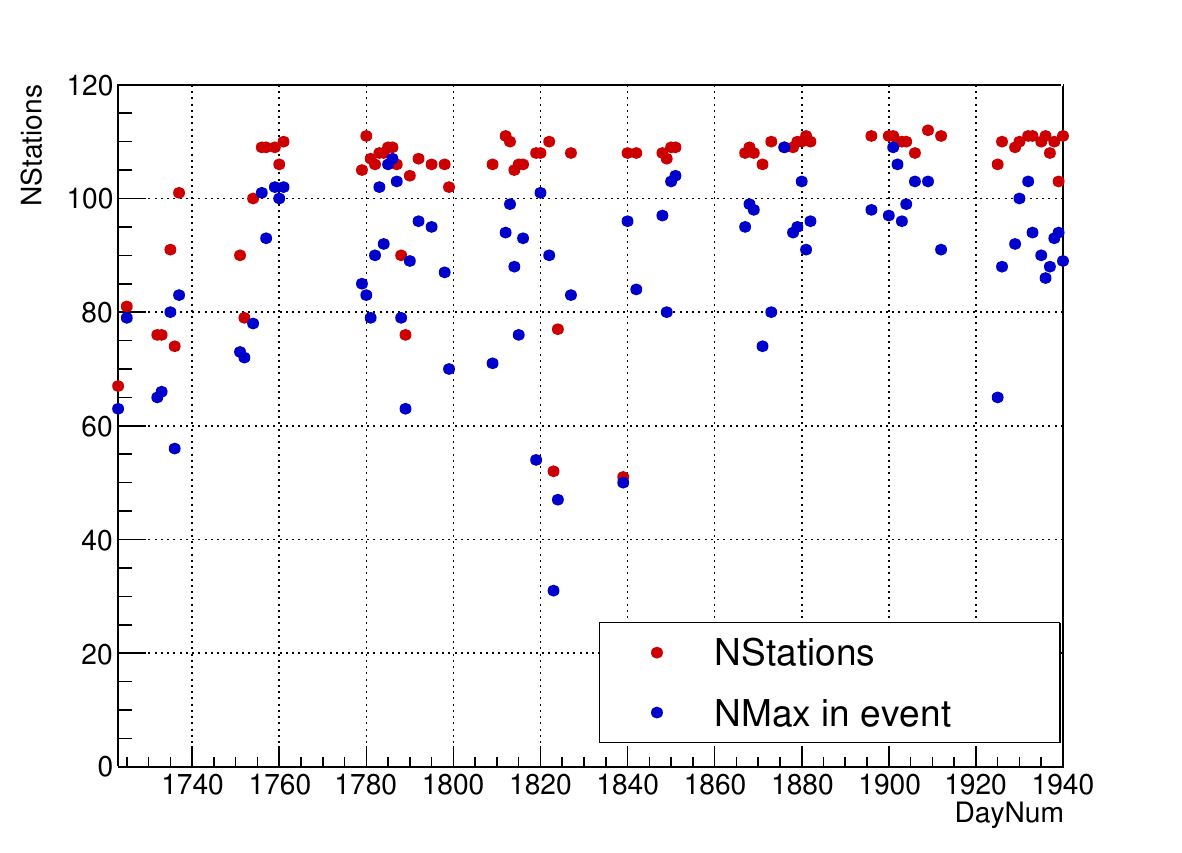}
\includegraphics[width=0.49\textwidth]{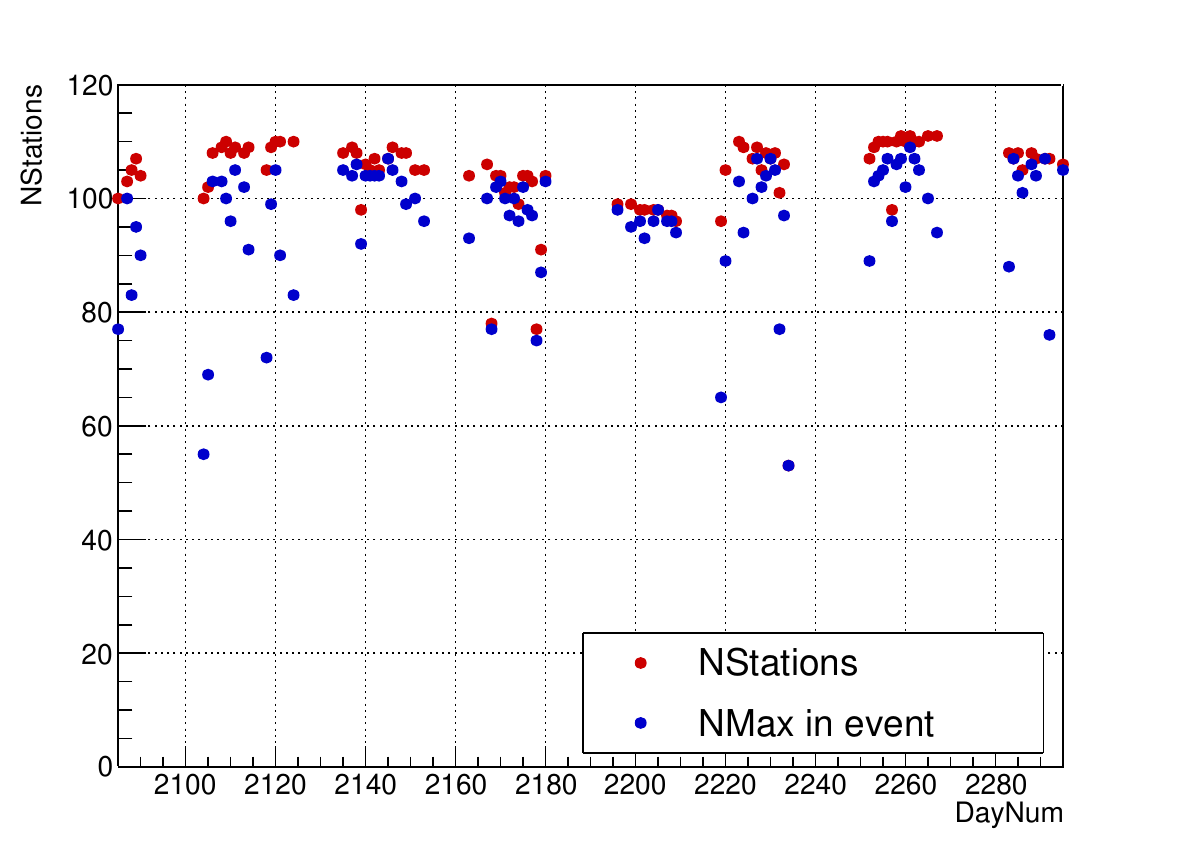}
\caption{Actual number of operating optical stations variations on individual nights of statistics collection. Time is measured in days counted consecutively from January 1, 2018. The parameter ``NStations'' shows the number of array stations triggered on a given night, and the parameter ``NMax in event'' shows the corresponding maximum number of stations triggered in a single event}\label{FigNMax}
\end{figure}

First of all, it is clear that the satellite's event area can depend on the number of actually operating optical stations in the HiSCORE array, out of a total of 120 stations in the array. Fig.~\ref{FigNMax} shows the actual number of operating stations (NStations parameter) for each measurement night for each of the 2022--2023 and 2023--2024 seasons. It is seen that, with the exception of the very beginning of the 2022--2023 season, the number of actually operating stations varied little, and was approximately 100--110. Therefore, we considered it possible to use a single setting of the filters for each season (separately), without changing them within a single season.

\begin{table}
\centering
\caption{The deviation of the satellite event times in each detected satellite passage from the optimal equidistant sequence of points and the optimal periods of pulses $\Delta T$ for each sattelite passage.\label{TabPassageFit}}
\begin{tabular}{|c|c|c|c|c|c|}
\hline
\multicolumn{3}{|c|}{2022--2023}             &  \multicolumn{3}{|c|}{2023--2024}\\
\hline
Date        & $\Delta T$ ($\mu$s) & $\sigma[t]$ ($\mu$s) & Date       & $\Delta T$ ($\mu$s) &$\sigma[t]$ ($\mu$s) \\
\hline
2022-10-03  & 5000.041            & 0.993               & 2023-09-18 & 5000.036            & 0.563 \\
2022-10-26  & 5000.038            & 0.652               & 2023-11-08 & 5000.040            & 1.579 \\
2022-12-30  & 5000.039            & 1.356               & 2023-11-15 & 5000.040            & 1.506 \\
2023-02-19  & 5000.038            & 1.492               & 2023-12-08 & 5000.044            & 0.028 \\
2023-03-19  & 5000.035            & 1.051               & 2023-12-15 & 5000.043            & 1.020 \\
-           & -                   & -                   & 2024-01-05 & 5000.042            & 1.289 \\
-           & -                   & -                   & 2024-03-03 & 5000.038            & 0.047 \\
-           & -                   & -                   & 2024-03-10 & 5000.036            & 1.024 \\
-           & -                   & -                   & 2024-03-12 & 5000.042            & 0.906 \\
-           & -                   & -                   & 2024-04-02 & 5000.039            & 1.638 \\
-           & -                   & -                   & 2024-04-09 & 5000.040            & 1.557 \\
\hline
\end{tabular}
\end{table}

To select satellite events for tuning the filters, we use a purely time-based selection method. This means the following.

First, we perform a preliminary selection of satellite events, which is easy to do since the satellite event chains are clearly visible above the chaotic distribution of EAS arrival directions in the sky. Then, using the preselected satellite events, an optimal equidistant sequence of events (time points) is fitted for each satellite passage separately. The quality of the fit is controlled. The deviation of the satellite event times in each detected sattelite passage from the optimal equidistant sequence of points and the optimal periods of pulses $\Delta T$ for each sattelite passage are shown in Tab.~\ref{TabPassageFit}. It can be seen that the satellite laser pulse frequency is approximately 20\,Hz and is maintained with a relative accuracy of about $10^{-4}$.

Then, within the time intervals containing the entire satellite passages, with a one-second additional margin at each end of the interval, all events that deviate in time from the optimal equidistant sequence points by no more than 20\,$\mu$s are selected. These events are considered to be the final sample of the satellite events, and exactly these points are shown in Fig.~\ref{FigSatt}. The diagrams in Fig.~\ref{FigSatt} clearly show that no extraneous EAS events were included in the sample: such events would not lie on the satellite trajectories. Importantly, this ``purely time-based'' method of selecting satellite events does not use any event filters that could distort the event statistics used to tune the filter thresholds to select the optical transient candidates.

\begin{figure}
\centering
\includegraphics[width=\textwidth]{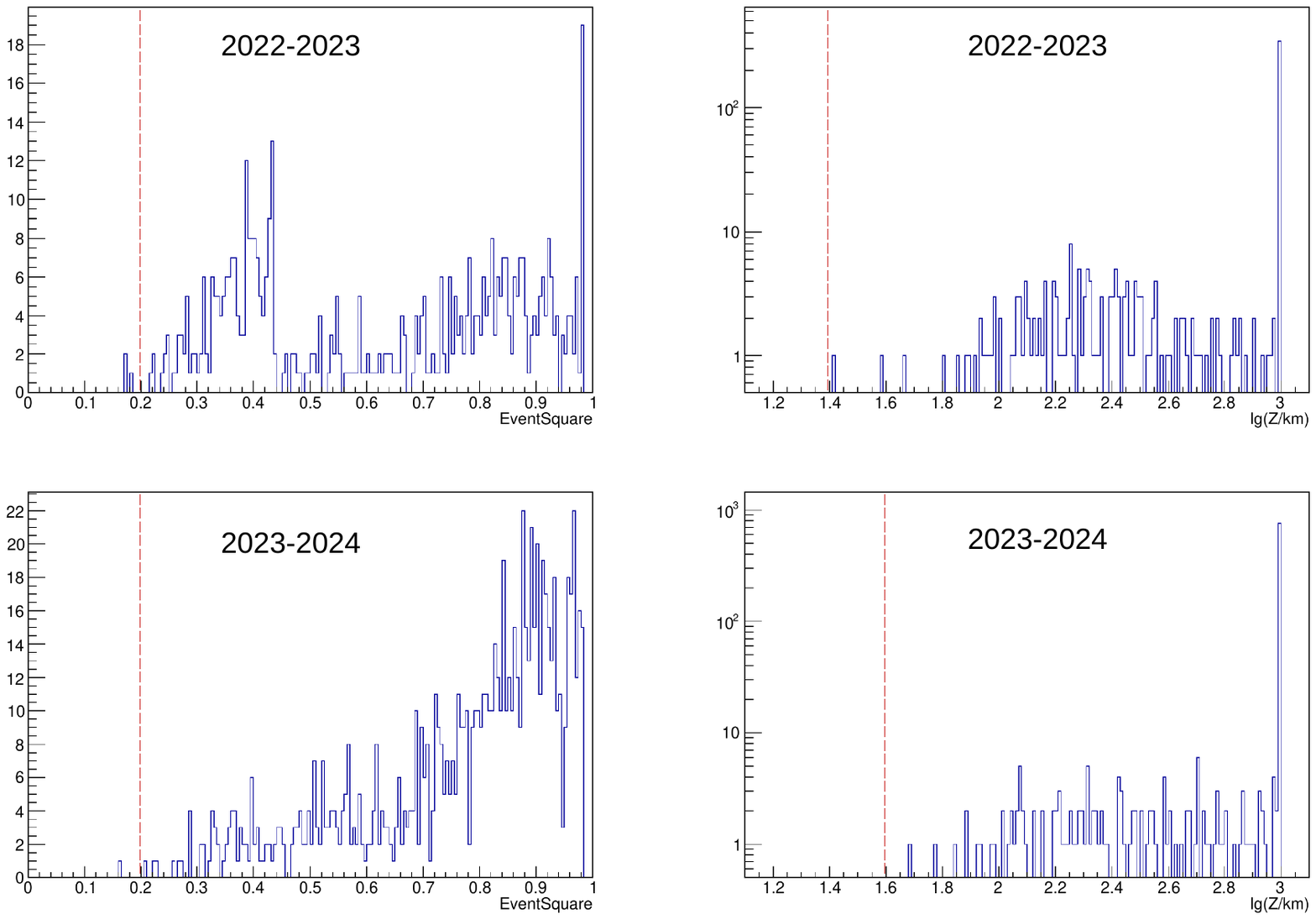}
\caption{Selection of the filter thresholds for searching candidates for distant astrophysical optical transients for the 2022--2023 and 2023--2024 seasons. The left panels show the event square distributions for all Daqi-1 events; the right panels show the distributions of the reconstructed spherical front center altitude $Z$ for the same events. The vertical dashed lines indicate the selected filter thresholds.}\label{FigSelection}
\end{figure}

Fig.~\ref{FigSelection} illustrates the selection of the filter thresholds for searching candidates for distant astrophysical optical transients for the 2022--2023 and 2023--2024 seasons. It is seen that the satellite's event parameters varied over a fairly wide range. This is due to two factors. First, the signal intensity varied significantly at different stages of the satellite's passage through the HiSCORE array's range of sight: from low at the beginning and end of the trajectory to high near the midpoint. Additionally, it sometimes varied within the satellite's trajectory due to the laser beam's diffraction pattern passing through the HiSCORE aperture. Second, the satellite's passages themselves were both ``strong'' and ``weak'', as could be seen by both the maximum amplitude of the detected signals and the duration of the passages. In any case, the detected events were associated with the passage of a nearly flat light front from a distant point source through the instrument, so the full range of \texttt{EventSquare} and $Z$ parameters variations must be taken into account.

The \texttt{EventSquare} and $Z$ thresholds for selecting optical transient candidates are shown in Fig.~\ref{FigSelection} by vertical dashed lines. The adopted values are $\mbox{\texttt{EventSquare}} = 0.2$ and $Z = 25$\,km for the season 2022--2023 and $\mbox{\texttt{EventSquare}} = 0.2$ and $Z = 40$\,km for the season 2023--2024. It is clear that this choice is very liberal. The selected thresholds reject only one or two weakest satellite events from the entire statistical dataset of several hundred events in each season. These weakest events are at the lower limit of visibility of the HiSCORE array. For some HiSCORE event to pass the optical transient candidate selection filter, both its \texttt{EventSquare} and $Z$ parameters must exceed the corresponding thresholds.

\begin{figure}
\centering
\includegraphics[width=\textwidth]{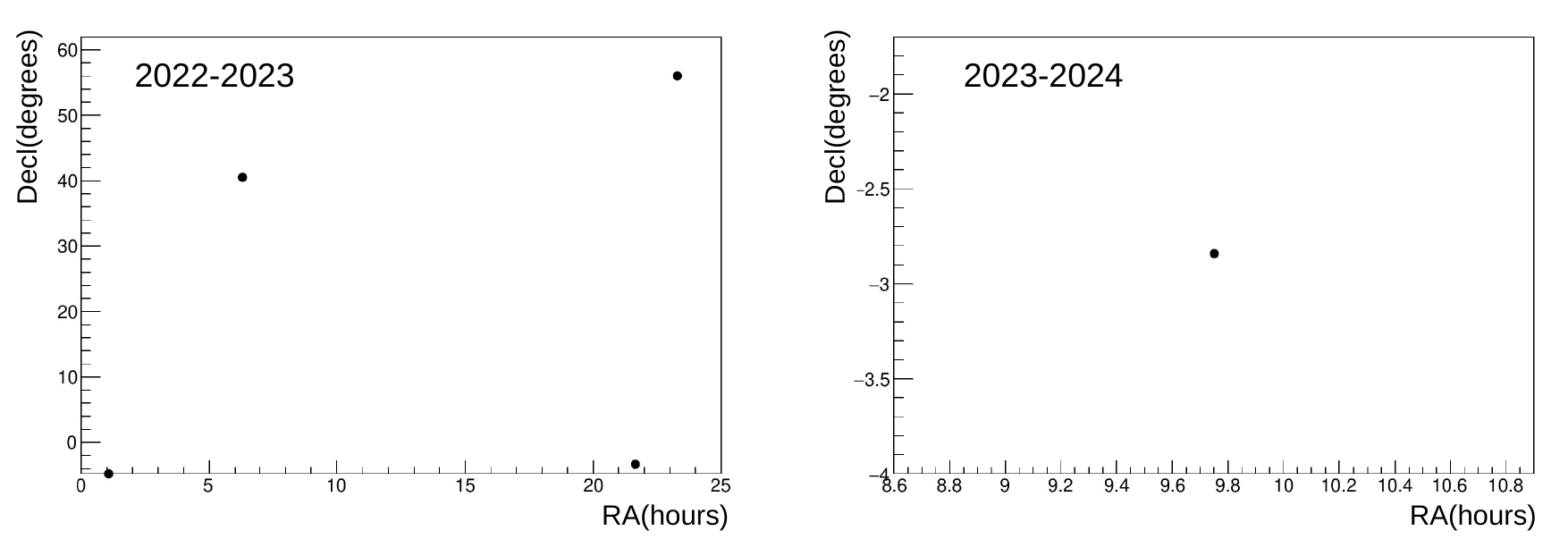}
\caption{The angular coordinates of optical transient candidates for the 2022--2023 and 2023--2024 seasons, passed the filtration by the \texttt{EventSquare} and $Z$ parameters. Equatorial coordinates are shown: declination (degrees) and right ascension (hours).}\label{FigCandidates}
\end{figure}

\begin{figure}
\centering
\includegraphics[width=0.49\textwidth]{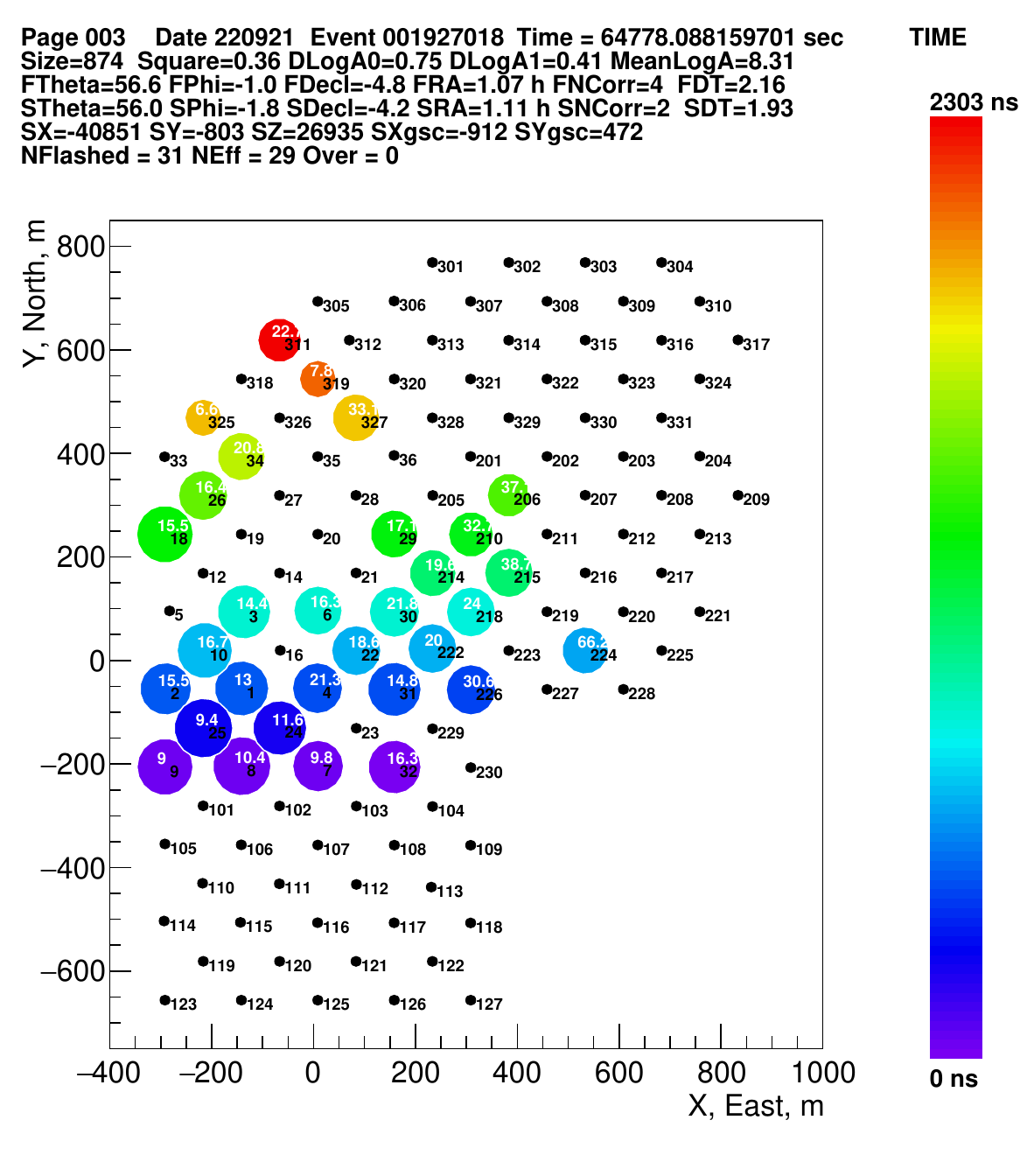}
\includegraphics[width=0.49\textwidth]{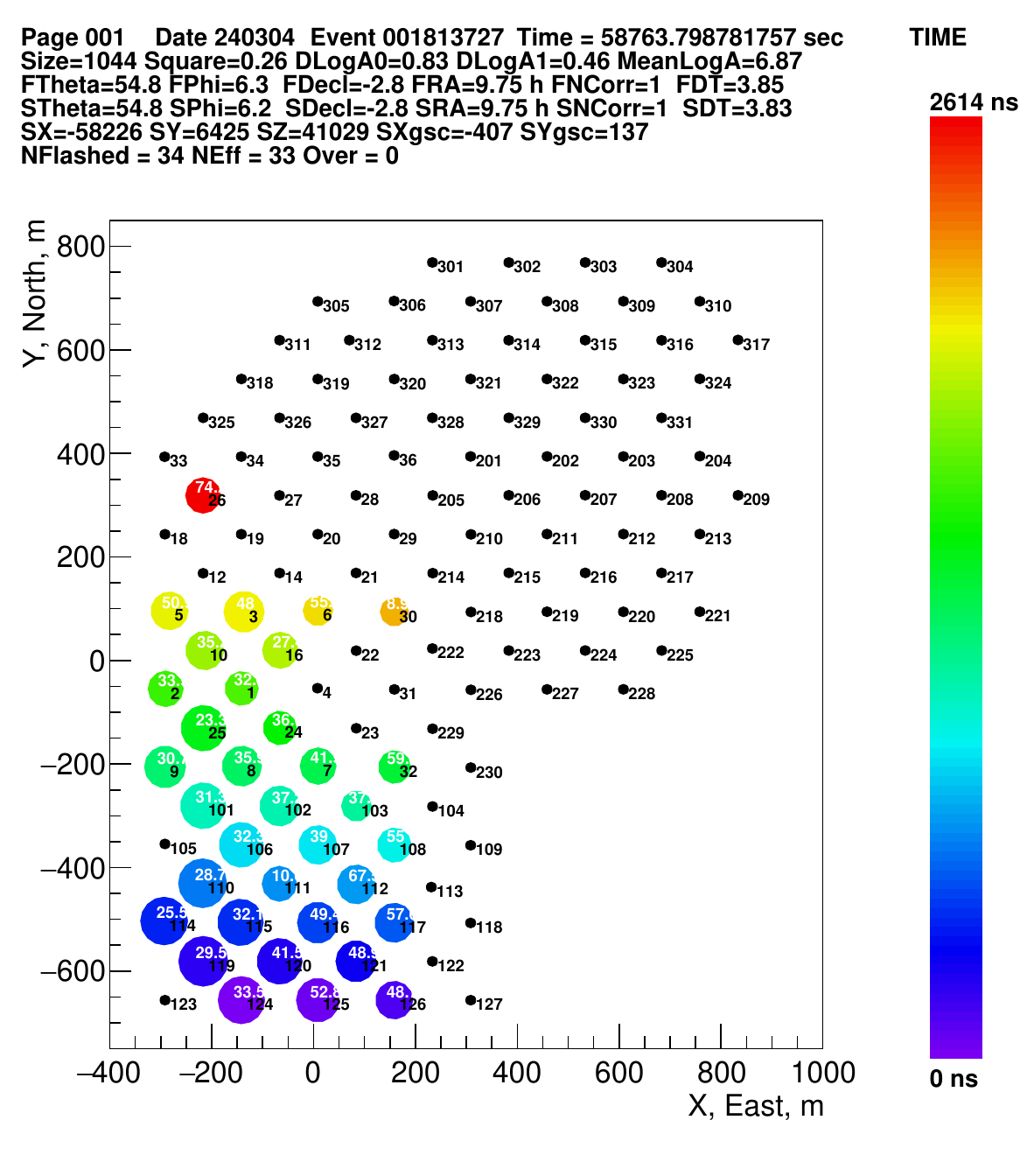}
\caption{Examples of events from the 2022-2023 (left) and 2023-2024 (right) seasons that passed the optical transient candidate filtering procedure.}\label{FigOpticatCand}
\end{figure}

The angular coordinates of optical transient candidates for the 2022--2023 and 2023--2024 seasons, passed the filtration by the \texttt{EventSquare} and $Z$ parameters, are shown in Fig.~\ref{FigCandidates}. It can be seen that four candidates passed the selection procedure in the 2022--2023 season, while only one passed in the 2023--2024 season. After automatic filtering, the selected events were visually inspected. This visual inspection revealed that none of the selected events could truly be considered to be optical transients. Fig.~\ref{FigOpticatCand} shows examples of events from the 2022--2023 and 2023--2024 seasons that passed the filtering procedure. For the 2022--2024 season, the most ``promising'' candidate is shown, while for the 2023--2024 season, the only candidate found is shown. In both cases, it is clear that, firstly, the array illumination is too concentrated in one local region of the array and, most importantly, the signal amplitudes of the optical stations exhibit noticeable non-uniformity, with large signals concentrated near a single point, which is simply the shower axis. That is, these are EAS events, not distant optical transient events.

Thus, no serious candidates for distant optical transients of astrophysical origin have been found in the 2022--2023 and 2023--2024 seasons.

\section{Search for narrow spatiotemporal groups of events (a signature of primordial black hole evaporation)}
\label{GG}

As can be seen from the section~\ref{OPT}, the first component of the multimesenger method for searching for black hole evaporation -- the search for optical flashes that could accompany black hole evaporation -- yielded a negative result. However, searching for narrow spatiotemporal groups of events, which could be another signature of black hole evaporation, \emph{has} also independent significance. This section presents the method and results of the search for narrow spatiotemporal groups of events with the TAIGA-HiSCORE Cherenkov array in the data of the 2022--2023 and 2023--2024 seasons of observation.

The definition of spatial (or more precisely, angular) proximity of events for searching for narrow groups is simple. Events are expected to arrive from exactly the same point in the sky during black hole evaporation, therefore the notion of ``spatial proximity'' of events will be defined simply by errors in determining angular coordinates.

The concept of temporal proximity of events is more complex. The energy threshold of the HiSCORE array is approximately 50\,TeV for detecting of gamma-rays. This value corresponds to events involving four optical stations. Since our data processing involves events of at least five stations, we will use a slightly higher threshold of 100\,TeV in our estimations.  Therefore detecting a narrow temporal group of gamma-quanta for HiSCORE means detecting at least two gamma-quanta with the energies above 100\,TeV within a time interval, which we will consider to be ``small'' in a certain sense. The question is the following. What is the time interval near the end of a black hole evaporation in which Hawking gammas with energies above 100\,TeV can be emitted? This time will define the scale of ``smallness'' of the time interval for defining the notion of a narrow temporal group of events.

\begin{figure}
\centering
\includegraphics[width=0.6\textwidth]{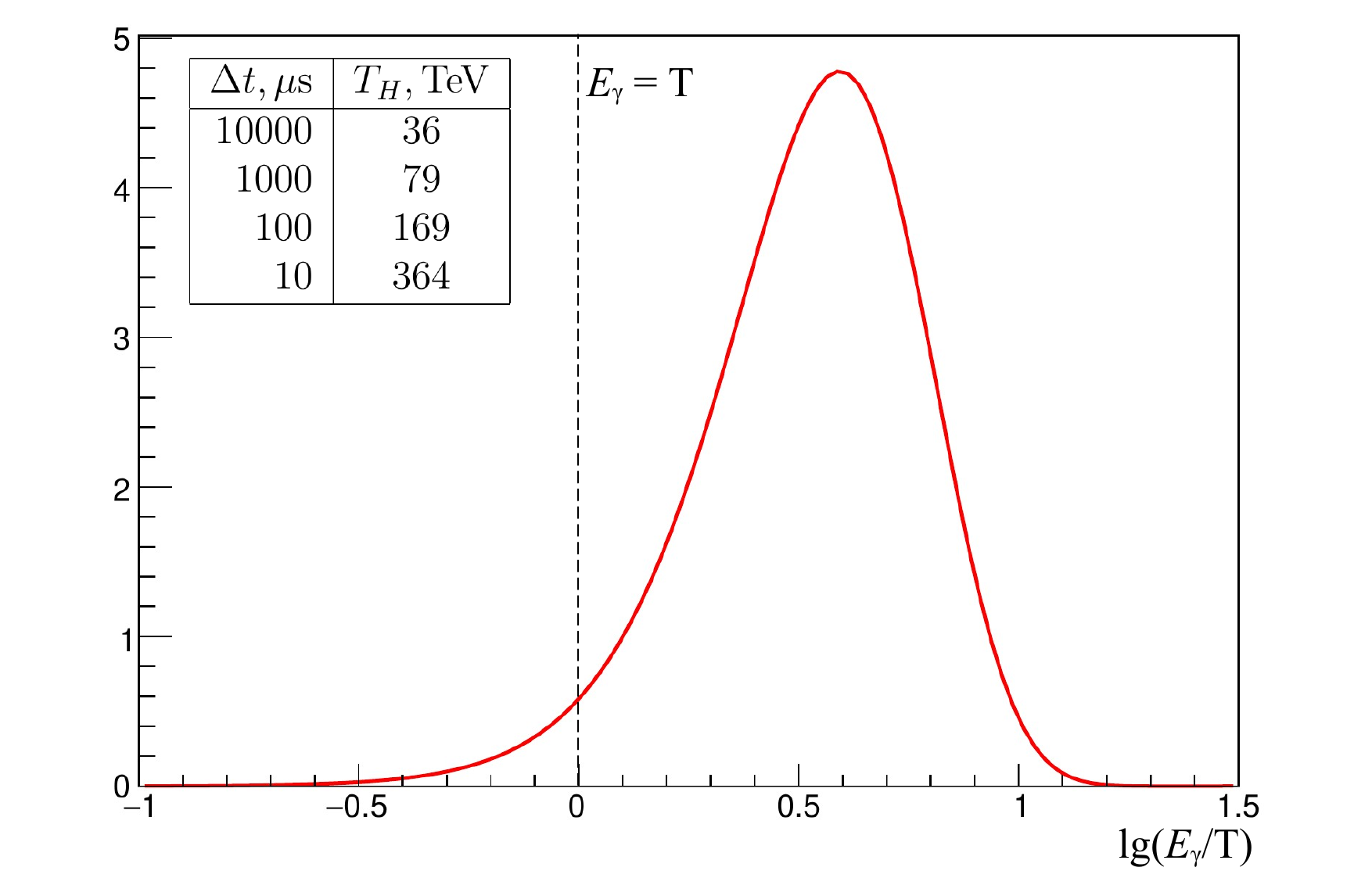}
\caption{Planck distribution of the energies $E_\gamma$ of Hawking radiation quanta for a black hole temperature $T_H$. The dependence of the Hawking temperature on the time $\Delta T$ until the end of the BH evaporation is also shown.}\label{FigPlanck}
\end{figure}

It is expected that the spectrum of Hawking gamma quanta is the black-body radiation Planck spectrum, the temperature of which is defined by the mass of the black hole. The Planck distribution of the energies $E_\gamma$ of Hawking gamma quanta for a Hawking temperature $T_H$ is shown in Fig.~\ref{FigPlanck}. The dependence of the Hawking temperature on the time $\Delta T$ until the end of the BH evaporation is also shown in a small table within the Fig.~\ref{FigPlanck}. It is evident from  Fig.~\ref{FigPlanck}  that Hawking gamma-rays with energies above 100\,TeV can be emmited intensively starting from a time of approximately 10\,ms until the end of BH evaporation, when the Hawking temperature is 36\,TeV. Gamma-rays with energies of 100\,TeV fall approximately at the maximum of the Planck distribution in this case. One might therefore think that short time intervals for searching for narrow groups of gamma-rays should be understood as time intervals of 10\,ms and less. However, it turns out that for a time interval of 10\,ms there will be a very large background of a chance EAS coincidences in the HiSCORE array (see below). We study in this work time intervals of 1000\,$\mu$s and below, down to fractions of a microsecond.

However, how can the HiSCORE array detect events with interval times between events shorter than 1000\,ns while events are formed using a 6500\,ns time window? We will demonstrate below how this problem can be solved, but before we begin to search for such narrow spatiotemporal event groups, we consider ``long'' inter-event time intervals, for which the maximum possible event duration of 6500\,ns can be considered to be short. We use 100\,$\mu$s and 1000\,$\mu$s windows as such time intervals.

First of all, we must understand the behavior of the EAS background in the formation of narrow groups of events by chance. For this end we simulate the EAS background in intervals of 100\,$\mu$s and 1000\,$\mu$s, separately for the full observing seasons of 2022--2023 and 2023--2024. Furthermore,  at this stage we consider time intervals between events greater than 7.5\,$\mu$s only, to avoid the complex problem of possible overlapping of events close in time.

\begin{figure}
\centering
\includegraphics[width=\textwidth]{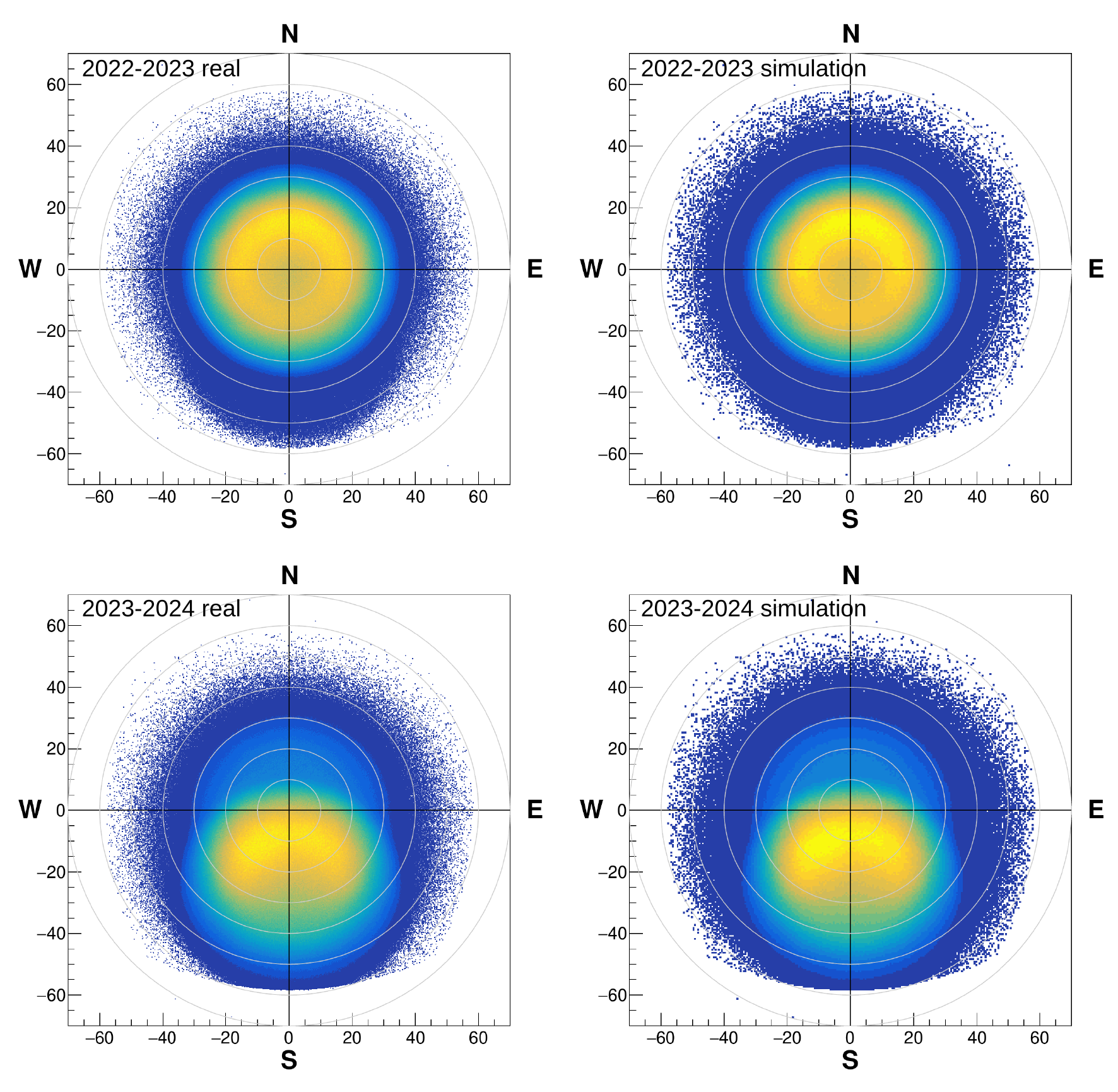}
\caption{Real measured arrival angle distributions of EAS events (left) and simulated angle distribution functions as histograms with $0.5^\mathrm{o}\times0.5^\mathrm{o}$ bins (right) for the 2022--2023 and 2023--2024 seasons. An orthogonal coordinate system with south-north and west-east axes is used, the angles on the axes are in degrees.}\label{FigSkyMap}
\end{figure}

To simulate the background of narrow groups of events, we first simulate the full event flow separately for each season.

For the event flow over time, we assume a Poisson process whose average rate is defined by the total exposure time per season and the total number of detected events with a multiplicity of at least 5 optical stations (see Section~\ref{DATA}). We assume a uniform Poisson process in the simulation, which, however, does not exactly correspond to reality, as the counting rate of the array could actually vary within certain limits, for example, due to the atmospheric conditions. In other words, we describe a Poisson process on average, and this approximation may introduce small systematic errors in the simulation.

The real measured arrival angle distributions of EAS events and simulated angle distribution functions as histograms with $0.5^\mathrm{o}\times0.5^\mathrm{o}$ bins for the 2022--2023 and 2023--2024 seasons are shown in Fig.~\ref{FigSkyMap}. It is seen that the distributions were different in different seasons of observation since the optical stations were directed to zenith in 2022--2023 but they were tilted at an angle of 30$^\mathrm{o}$ from the zenith to the south direction in 2023--2024.

The notion of spatial (angular) proximity of events is defined exactly as follows. When processing the real data from the HiSCORE array, the EAS arrival direction is determined for each detected event by approximating the light front of this event with a spherical front. For each such approximation, the errors (in terms of standard deviations) in the determined angular coordinates are calculated separately. The errors in the angular coordinates are associated with random errors in measuring the trigger times of the optical stations. These errors is known from the test events obtained by flashing a photodiode mounted on a drone that performed several test flights over the HiSCORE array at an altitude of 400--500\,m. Since the light front of a point-like photodiode is exactly spherical, the algorithm for approximating the light front with a spherical front should provide an accurate approximation up to the random errors in the trigger times of the optical stations. The RMS error in the time measurements can be determined from the deviation of the approximation from the actual measured trigger times of the optical stations. It turned out to be $\sigma[t] = 0.63$\,ns.

\begin{figure}
\centering
\includegraphics[width=0.6\textwidth]{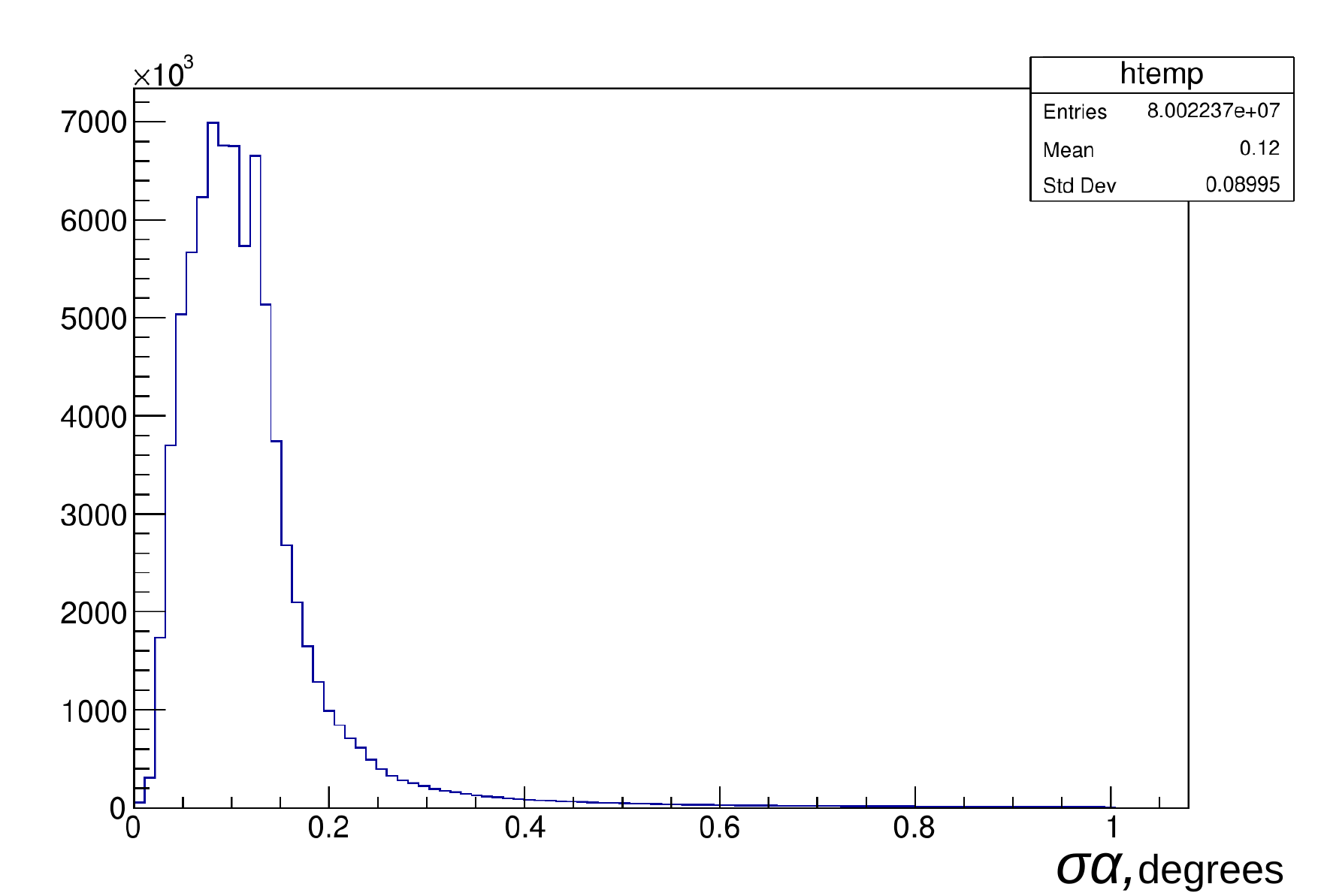}
\caption{Distribution of the angular error $\sigma[\alpha]$ of the  algorithm of event direction reconstruction using spherical front fit for all statistics of the 2022--2023 season.}\label{FigDirError}
\end{figure}

For each event, two angular errors are calculated separately, corresponding to two mutually orthogonal angular directions centered at the center of the event. These errors are not necessarily identical, but in the algorithm for determining the spatial proximity of events, we used the ``combined error'' $\sigma[\alpha]$, which is defined as the square root sum of the squares of the individual errors, instead of these two errors individually. Below we will call such a combined angle error simply ``an error''.

The distribution of the angular errors $\sigma[\alpha]$ of the  algorithm of event direction reconstruction using spherical front fit for all statistics of the 2022--2023 season is shown in Fig.~\ref{FigDirError}. The distribution for the 2023--2024 season is not significantly different from the distribution for the 2022--2023 season. It is seen that the distribution is asymmetric. The long tail on the right is associated with rare statistical anomalies in the form of EAS in the HiSCORE array. The majority of regular events have an angular error of no more than 0.2$^\mathrm{o}$, and only among these regular events a search for narrow spatiotemporal groups of events was carried out.

To check the spatial proximity of a group of events that are already close in time, the following algorithm is used. A group of events is considered to be a group of spatially close  events if the following condition is met for each event. The angular center of weight of the group is calculated with weights proportional to $1/\sigma[\alpha_i]$ ($i$ is the number of the event in the group), and if the angular distance of each event from the center of weight does not exceed $1.5\times\sigma[\alpha_i]$, then such a group is considered to be a narrow spatial group. For example, if a group contains only two events with identical angular errors $\sigma[\alpha_1] = \sigma[\alpha_2] = \sigma[\alpha]$, then the events is a narrow spatial pair if the angular distance between the events is no more than $3\times\sigma[\alpha]$.

An event group is considered temporally narrow in the first series of our tests if the time interval from the first event to the last one does not exceed the specified time window (100\,$\mu$s or 1000\,$\mu$s) and all the time intervals between successive events in the group exceed 7.5\,$\mu$s (see the explanation above for the reason). Finally, an event group forms a narrow spatiotemporal group if it is both spatially narrow and temporally narrow.

When simulating the background of narrow groups of events, the arrival direction is simulated for each event according to the probability distributions in Fig.~\ref{FigSkyMap}. Then, for each event independently, the $\sigma[\alpha]$ error value is simulated according to the probability distribution in Fig.~\ref{FigDirError}. The time intervals between successive events are simulated according to an exponential distribution, which follows from the Poisson nature of the process. The selection of narrow spatiotemporal groups of simulated events is carried out by using exactly the same described above algorithm as when analyzing the statistics of the real data.

\begin{table}
\centering
\caption{The number of close spatiotemporal pairs of events for inter-event time intervals of $7.5\,\mu\mathrm{s} < \Delta t < 100\,\mu\mathrm{s}$ and $7.5\,\mu\mathrm{s} < \Delta t < 1000\,\mu\mathrm{s}$ based on real data from the 2022--2023 and 2023--2024 seasons and according to the simulations.\label{TabLongPares}}
\begin{tabular}{|c|c|c|c|c|c|}
\hline
\multicolumn{6}{|c|}{$7.5\,\mu\mathrm{s} < \Delta t < 100\,\mu\mathrm{s}$}\\
\hline
Season        & $N$ (data) & $\sigma[N(\mathrm{data})]$ & $N$ (simul) & $\sigma[N(\mathrm{simul})]$  & difference \\
\hline
2022--2023    & 31         & 6                 & 23.50       & 0.15               & $+1.3\sigma$ \\
2023--2024    & 27         & 5                 & 29.01       & 0.17               & $-0.4\sigma$ \\
\hline
\multicolumn{6}{|c|}{$7.5\,\mu\mathrm{s} < \Delta t < 1000\,\mu\mathrm{s}$}\\
\hline
Season        & $N$ (data) & $\sigma[N(\mathrm{data})]$ & $N$ (simul) & $\sigma[N(\mathrm{simul)}]$  & difference \\
\hline
2022--2023    & 338        & 18                & 251.0       & 0.5                & $+4.8\sigma$ \\
2023--2024    & 295        & 17                & 306.4       & 0.6                & $-0.7\sigma$ \\
\hline
\end{tabular}
\end{table}

The number of close spatiotemporal pairs of events for inter-event time intervals of $7.5\,\mu\mathrm{s} < \Delta t < 100\,\mu\mathrm{s}$ and $7.5\,\mu\mathrm{s} < \Delta t < 1000\,\mu\mathrm{s}$ based on real data from the 2022--2023 and 2023--2024 seasons and according to the simulations are shown in Tab.~\ref{TabLongPares}.
No narrow triple group of events was found in either real data or in simulation results.

\begin{figure}
\centering
\includegraphics[width=0.49\textwidth]{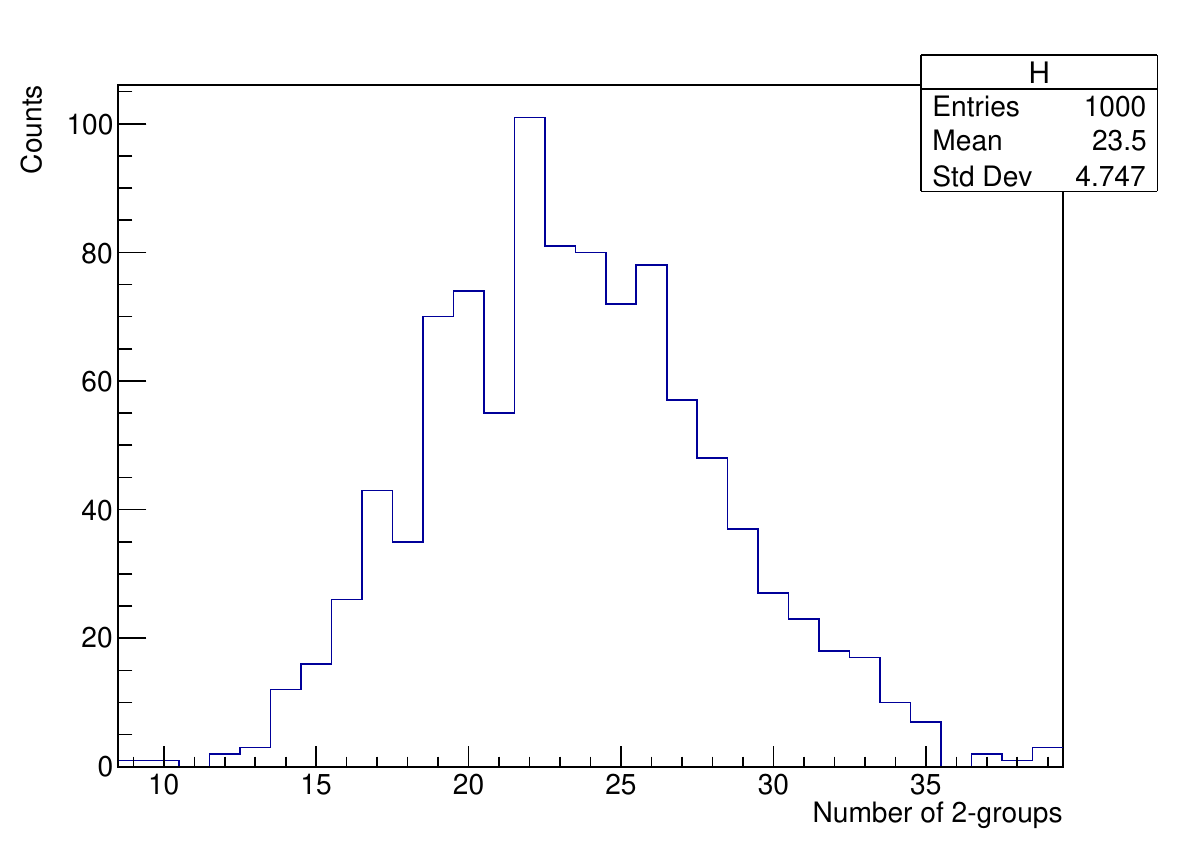}
\includegraphics[width=0.49\textwidth]{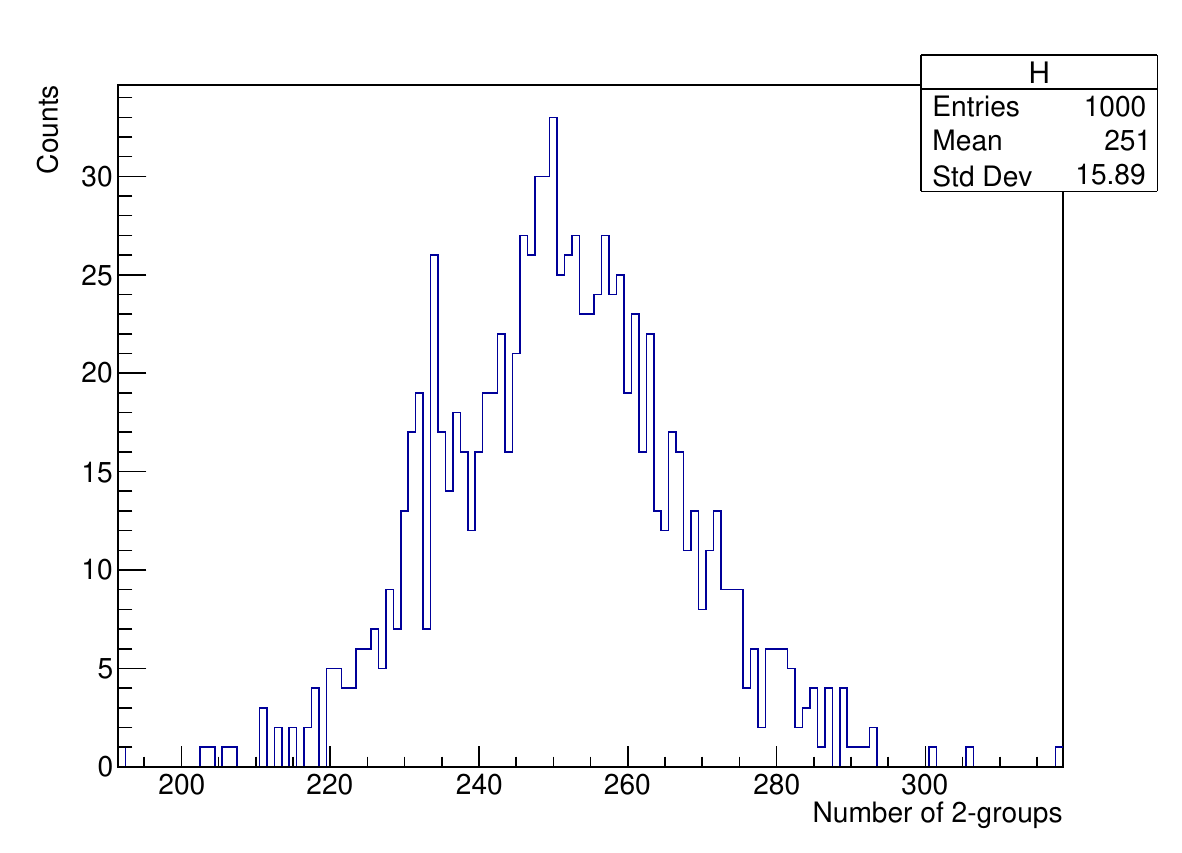}
\caption{Distributions of the number of narrow pairs of events in the MC-simulation in the 2022--2023 season for the intervals $7.5\,\mu\mathrm{s} < \Delta t < 100\,\mu\mathrm{s}$ (left panel) and $7.5\,\mu\mathrm{s} < \Delta t < 1000\,\mu\mathrm{s}$ (right panel).}\label{FigLongWinSimul}
\end{figure}

In Monte Carlo simulations, 1000 tests were performed for each season and for each of the time windows $7.5\,\mu\mathrm{s} < \Delta t < 100\,\mu\mathrm{s}$ and $7.5\,\mu\mathrm{s} < \Delta t < 1000\,\mu\mathrm{s}$. Each test corresponds to one full exposure of the HiSCORE instrument over the entire season. The distributions of the number of narrow pairs of events in the MC-simulation in the 2022--2023 season for the intervals $7.5\,\mu\mathrm{s} < \Delta t < 100\,\mu\mathrm{s}$ and $7.5\,\mu\mathrm{s} < \Delta t < 1000\,\mu\mathrm{s}$ are shown in Fig.~\ref{FigLongWinSimul}. The distributions for the 2023--2024 season look similar. As can be seen from Fig.~\ref{FigLongWinSimul}, the standard deviation of the distributions is related to the mean values as $\sigma[N] = \sqrt{N}$. Therefore, to estimate the statistical uncertainty in the number of narrow pairs of events found in real observations, the estimate $\sigma[N] = \sqrt{N}$ was used as well. The corresponding RMS deviations are given in the column ``$\sigma[N(\mathrm{data})]$'' of Tab.~\ref{TabLongPares}. The column ``$\sigma[N (\mathrm{simul})]$'' shows the errors of the mean value for the distributions like in Fig.~\ref{FigLongWinSimul} for the 2022--2023 and 2023--2024 seasons. It is clear that the error of the simulated mean value is much less than the experimental error and can be neglected when comparing the simulation and the experiment.

The ``difference'' column in Tab.~\ref{TabLongPares} shows the difference between the simulated number of narrow event pairs and those found in the real data, expressed as a number of standard deviations. It is seen that for narrow event pairs within intervals of $7.5\,\mu\mathrm{s} < \Delta t < 100\,\mu\mathrm{s}$, there are no statistically significant differences between the experimental values and the simulation. For narrow event pairs within intervals of $7.5\,\mu\mathrm{s} < \Delta t < 1000\,\mu\mathrm{s}$, the found number of narrow pairs for the 2022--2023 season exceeds the simulation by $+4.8\sigma$, but for the 2023--2024 season it has the opposite sign and is $-0.7\sigma$. The statistically significant deviation of $+4.8\sigma$ in 2022--2023 season could well be explained by an unaccounted of a factor related to the variability of the event counting rate in the real experiment, which was approximated only on average, by a homogeneous Poisson process in the simulation, as has been noted above. Since the deviation in 2023--2024 season has the opposite sign, then one can conclude that there is no pronounced systematics in deviation of the data from the simulation in the case $7.5\,\mu\mathrm{s} < \Delta t < 1000\,\mu\mathrm{s}$ either.

Thus, a comparison of the simulated background value and the real data shows that there is no significant excess of the number of close spatiotemporal pairs of events over background expectations for the ``big'' intervals $7.5\,\mu\mathrm{s} < \Delta t < 100\,\mu\mathrm{s}$ and $7.5\,\mu\mathrm{s} < \Delta t < 1000\,\mu\mathrm{s}$.

The above analysis show that the relative accuracy of the background estimate is, in any case, no worse than approximately 25\%. This accuracy is sufficient to begin searching for narrow spatiotemporal groups of events with time intervals between events $0 < \Delta t < 7.5\,\mu\mathrm{s}$.

Since the time window for generating a single HiSCORE event from triggers of individual optical stations is 6.5\,$\mu$s, particle arrivals separated by less than 6.5\,$\mu$s time interval cannot be detected by HiSCORE as separate events. However, if two (or more) EASs arrive at the HiSCORE array with a time interval of less than 6.5\,$\mu$s and the spatial size of the EASs at ground level is not too large, the HiSCORE event generation algorithm can combine these individual EASs into a single event, which could be distinguished by its complex spatial structure. The event will not have the form of a single ``spot'', as in Fig.~\ref{FigGenericEAS} on the left, but will instead have the appearance of two (or more) clearly distinguishable ``spots'' -- subevents. To verify that such a complex event represents a narrow spatiotemporal group, each such subevent should be processed separately by the shower's arrival direction reconstruction algorithm. If both such subevents point to the same sky point (within the accuracy of the direction reconstruction errors, as described above), then such a complex event can be considered to be a candidate for a narrow spatiotemporal pair (or group). The temporal narrowness of such a pair is automatically ensured by the fact that the subevents are included in a single full HiSCORE event with a duration of no more than 6.5\,$\mu$s. Details of the temporal structure of such multiple events will allow us to reconstruct the exact difference in the arrival times of the component events (see details below).

The algorithm for searching for multiple events -- candidates for narrow spatiotemporal pairs -- is as follows. Among all HiSCORE events, we search for those in which the event structure at ground level resembles two (or more) disconnected spots -- subevents. There are a great many such complex events in the HiSCORE data. For each subevent, the shower direction is reconstructed, and if the subevent axes point to the same point in the sky, the event is considered to have passed the primary filter and becomes a candidate for narrow spatiotemporal pairs. However, such events also turn out to be very numerous. The origin of such events is that an EAS from a single particle can be detected as two spatially disconnected spots simply for a chance reasons -- due to the spatial inhomogeneity of the shower or because some optical stations failed to operate due to electronic instability. This also happens, albeit rarely. Since both subevents in such cases are part of the same EAS, the reconstructed directions of their axes naturally point to the same point in the sky.

\begin{figure}
\centering
\includegraphics[width=\textwidth]{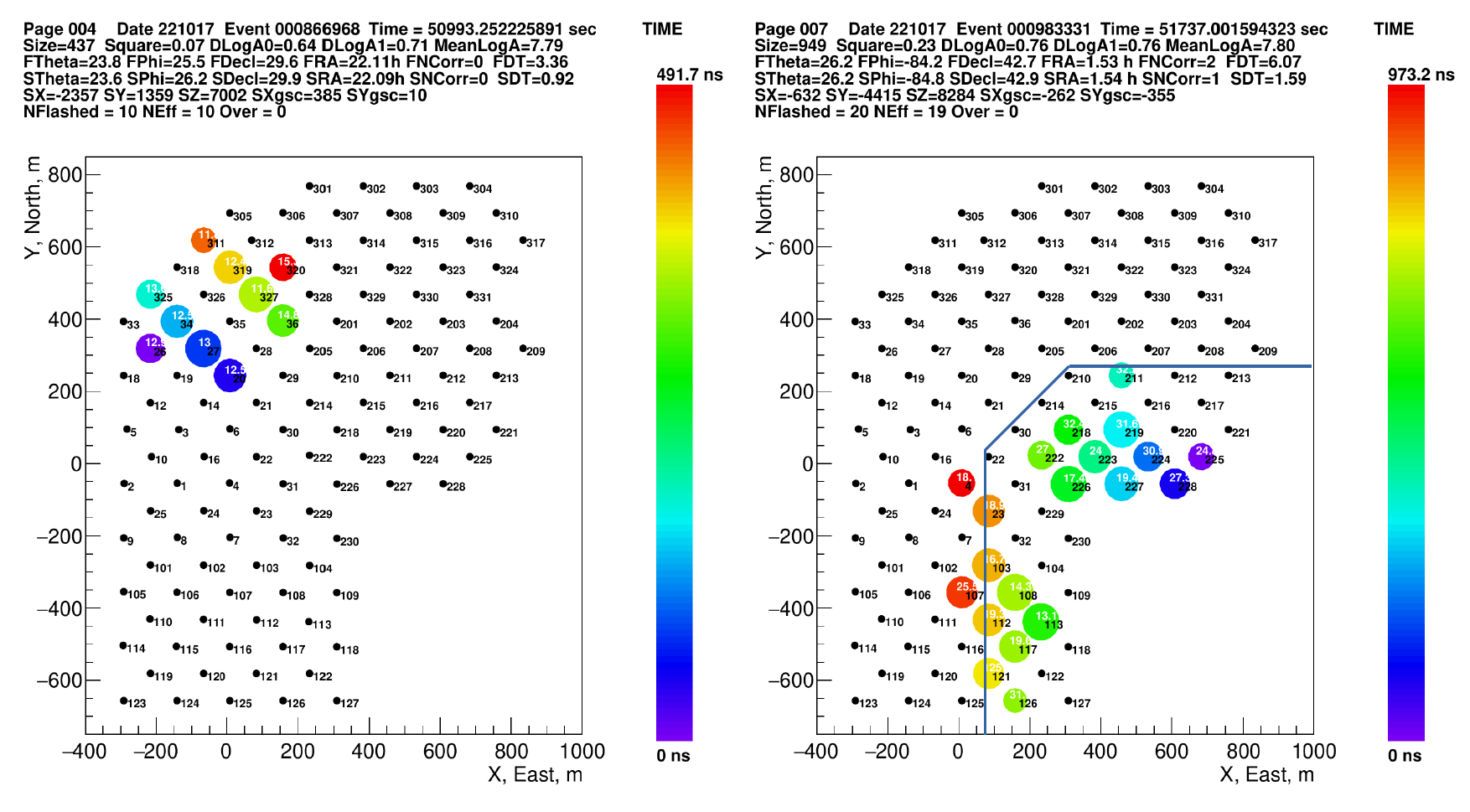}
\caption{Two examples of ``incorrect'' narrow spatiotemporal pairs of events.}\label{FigInvalidPairs}
\end{figure}

Fig.~\ref{FigInvalidPairs} shows two typical examples of ``incorrect'' narrow spatiotemporal pairs of events. The event in the left panel of Fig.~\ref{FigInvalidPairs} is cut into two disconnected parts by a line of stations numbered 28, 35, 328, and 318. This pattern of event discontinuity is repeated several times in the data of 2022--2023 and 2023--2024 seasons. This is apparently due to this line of stations exhibiting some additional instability compare to the other optical stations. Similar events, but less common, is also encountered in other parts of the HiSCORE array. ``Narrow pairs'' of this type should not be considered as real candidates for narrow pairs. The event in the right panel of Fig.~\ref{FigInvalidPairs} also represents a widespread pattern of ``incorrect'' narrow pairs. The origin of events of this type is related to the fact that the lower right corner of the HiSCORE array is ``cut out''. There are a large number of events in which the shower axis falls into this cut-out corner, where the shower cannot be detected, but the EAS wings are detected by optical stations located near this cut-out corner. Since statistical fluctuations in the Cherenkov light flux in EAS wings are large, discontinuities in the shower structure can often arise here naturally. Events of this type should also not be considered as candidates for narrow pairs.

To avoid ``false narrow pairs'', additional conditions were imposed when selecting candidates for narrow pairs. First, the distance in the HiSCORE array plane between the nearest optical stations of different components must exceed 250\,m. This avoids incorrect events, as in Fig.~\ref{FigInvalidPairs}, left panel. Second, the event must not be a ``lower right corner event'' of the HiSCORE array, as in Fig.~\ref{FigInvalidPairs} on the right. The precise formulation of this condition was as follows. If half or more of the optical stations in each subevent are within the contour shown in Fig.~\ref{FigInvalidPairs} (right panel) by the solid line, then the event is considered to be a ``corner event'' and rejected. Another requirement when filtering candidates for narrow pairs is that the size of each subevent must be at least five triggered stations. This requirement is simply related to the specifics of the current version of the shower arrival direction reconstruction algorithm using a spherical front, which requires at least five stations in one event to work. Another one additional criterion was the condition that the time interval between subevents of one event should not be more than 6\,$\mu$s, in order to avoid problems with possible sub-events going beyond the maximum possible duration of the event, which is 6.5\,$\mu$s.

In order to understand the results of the search for narrow spatiotemporal groups of events in the real data, it is necessary to estimate the expected background of accidental EAS coincidences, as we have already done above for the time windows $7.5\,\mu\mathrm{s} < \Delta t < 100\,\mu\mathrm{s}$ and $7.5\,\mu\mathrm{s} < \Delta t < 1000\,\mu\mathrm{s}$. Now it is necessary in the simulation to choose the time window $0 < \Delta t < 6\,\mu\mathrm{s}$, but this is not enough. It must also be taken into account that some of the real narrow spatiotemporal pairs will be discarded because the subevents accidentally come too spatially close to each other in the plane of the HiSCORE array, therefore either simply overlapping one another, or not satisfying the criterion of ``more than 250\,m between subevents''. Subevents may also randomly fall into the ``lower right corner'' region and will also be rejected by the selection criteria. Therefore, the procedure for selecting candidates for narrow pairs has a finite efficiency which is less than unit. A Monte Carlo simulation using a real, known event size distribution and assuming a random shape of the EAS events in the array plane yields an efficiency of 0.527 ($10^6$ Monte Carlo trials).

\begin{figure}
\centering
\includegraphics[width=\textwidth]{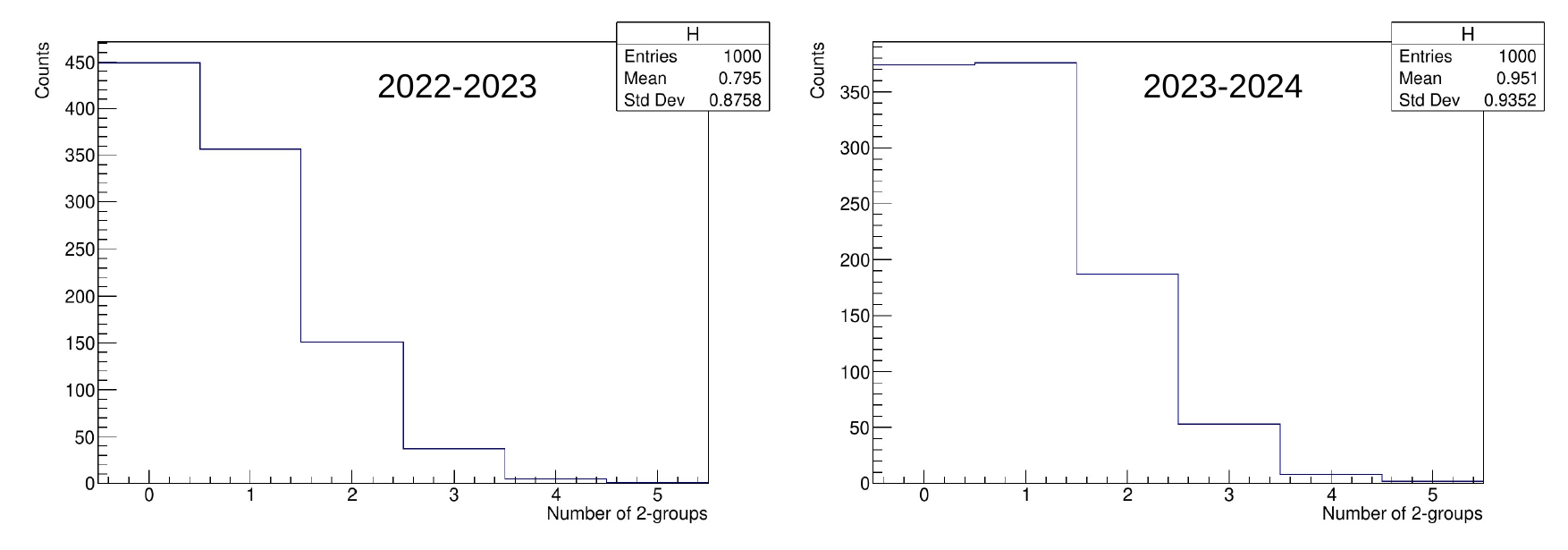}
\caption{Distributions of the number of backgound narrow spatiotemoral pairs of events in the MC-simulation for the time intervals $0 < \Delta t < 6\,\mu\mathrm{s}$ with accounting for the efficiency of registration of 0.527 (see text) for the seasons 2022--2023 and 2023--2024.}\label{FigMC0006}
\end{figure}

\begin{figure}
\centering
\includegraphics[width=0.49\textwidth]{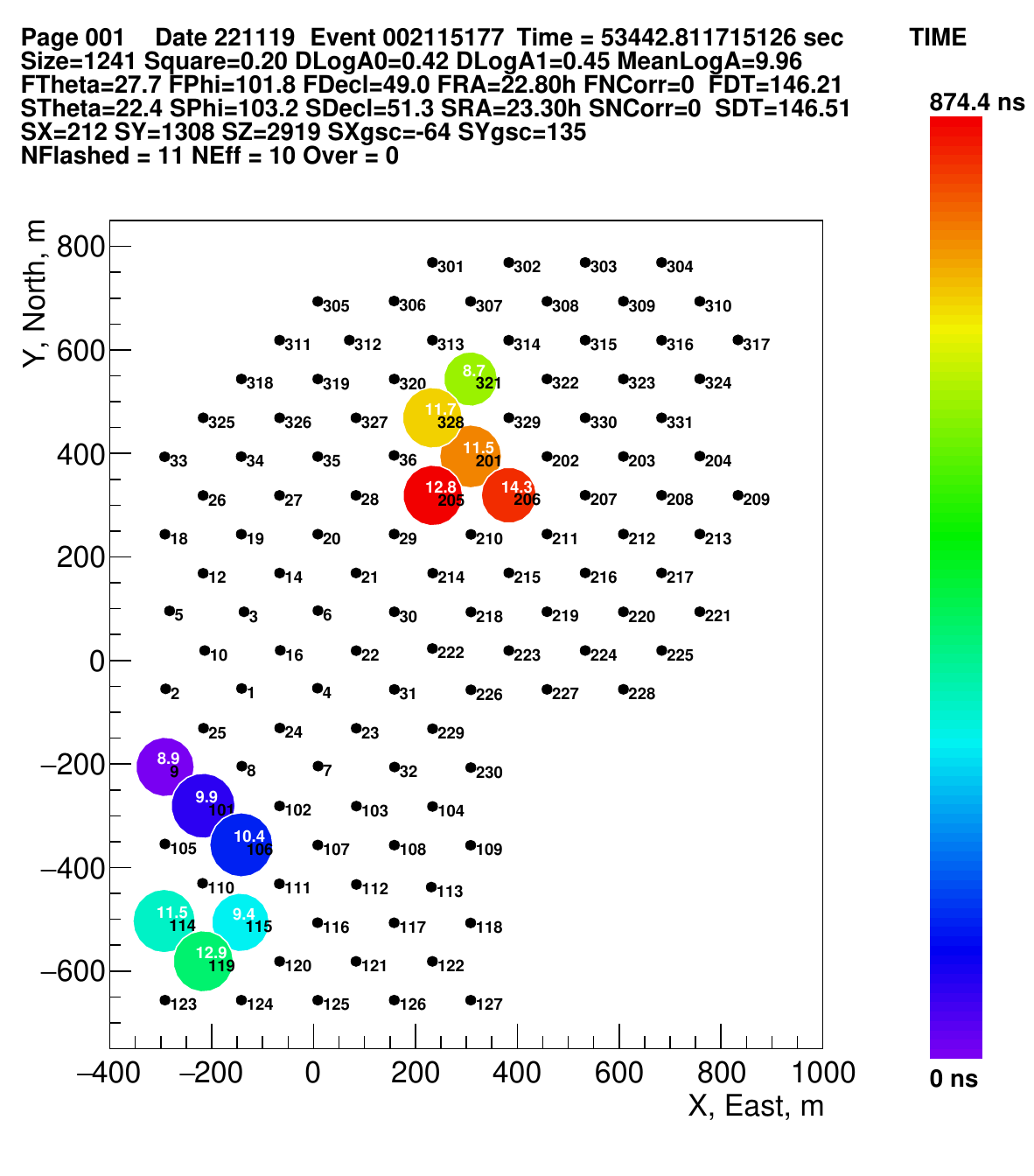}
\includegraphics[width=0.49\textwidth]{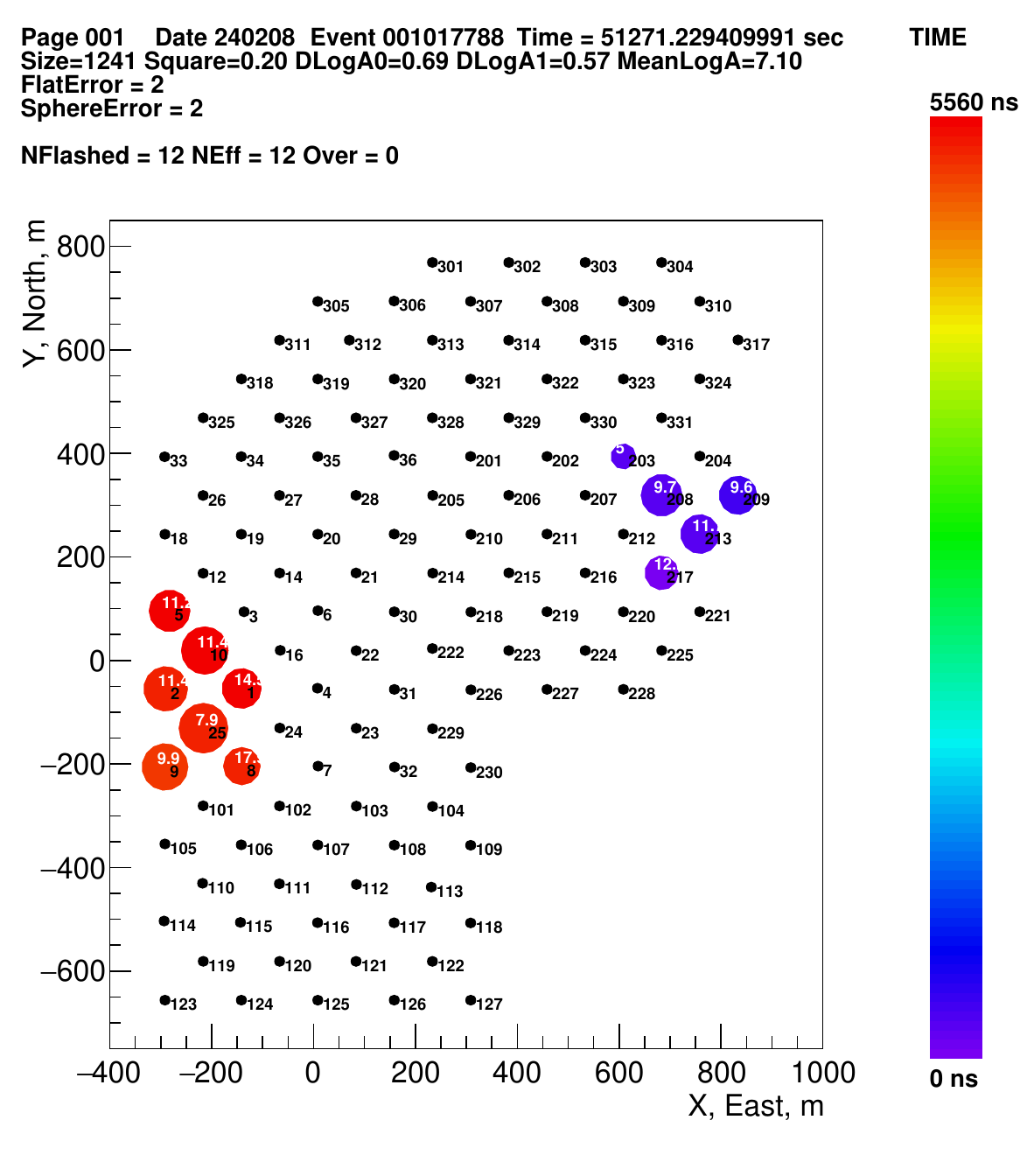}
\caption{Narrow spatiotemporal pairs of events found in the HiSCORE array data in the 2022-2023 (left) and 2023-2024 (right) observing seasons.}\label{FigEvent}
\end{figure}

The distributions of the number of backgound narrow spatiotemoral pairs of events in the MC-simulation for the time intervals $0 < \Delta t < 6\,\mu\mathrm{s}$ with accounting for the efficiency of registration for the seasons 2022--2023 and 2023--2024 are shown in Fig.~\ref{FigMC0006}. It can be seen that the expected value of the accidental coincidence background is approximately one narrow pair of events for each of the 2022--2023 and 2023--2024 seasons, with highly probable values from 0 to 3. In the real HiSCORE data, all filters passed only two narrow spatiotemporal pairs, exactly one in each of the 2022--2023 and 2023--2024 observing seasons. Both these compex events are shown in Fig.~\ref{FigEvent}.

\begin{table}
\centering
\caption{Parameters of narrow spatiotemporal pairs of events found in HiSCORE data (see text).\label{TabEvents}}
\begin{tabular}{|c|c|c|c|c|c|}
\hline
No & Season & $\Delta\alpha$ & $\Delta t$,\,ns & $\Delta t_{\mathrm{Corrected}}$,\,ns & $\sigma(\Delta t_{\mathrm{Corrected}})$,\,ns \\
\hline
1 & 2022--2023 & 0.18$^\circ$ &  612 & 1585.6 & 1.8 \\
2 & 2023--2024 & 0.10$^\circ$ & 5256 & 6563.9 & 1.5 \\
\hline
\end{tabular}
\end{table}

The first issue to note is that the number of detected narrow pairs is completely consistent with the background expectations. There are no indications of background exceedances and, consequently, the presence of astrophysical sources for narrow pairs of events. Some parameters of the detected narrow spatiotemporal pairs of events are given in Table.~\ref{TabEvents}. In the table, $\Delta\alpha$ denotes the angular separation between the detected components of narrow pairs, $\Delta t$ is the time interval between the triggers of subevents in the narrow pairs.

\begin{figure}
\centering
\includegraphics[width=\textwidth]{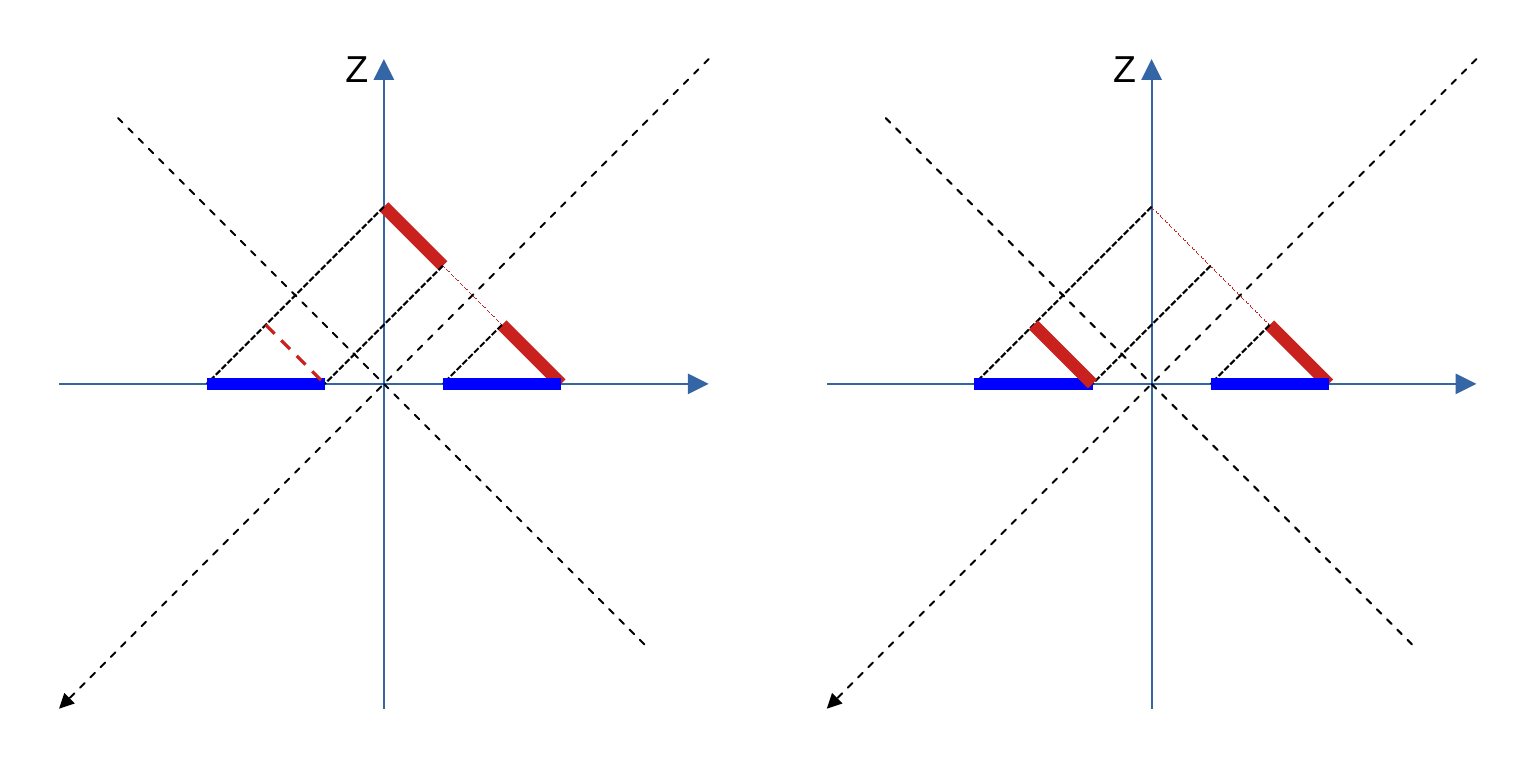}
\caption{Narrow spatiotemporal pairs of events found in the HiSCORE array data in the 2022-2023 (left) and 2023-2024 (right) observing seasons.}\label{FigDeltaTScheme}
\end{figure}

It should be noted, however, that the time interval between subevent triggers $\Delta t$ does not correctly represent the true difference in subevent arrival times. Fig.~\ref{FigDeltaTScheme} illustrates the geometry of the arrival of Cherenkov light fronts of a narrow spatiotemporal pair of events onto the HiSCORE array plane. In Fig.~\ref{FigDeltaTScheme}, the thick red inclined lines represent the positions of the light fronts of the subevents of the narrow pair at a certain moment in time, and the thick blue horizontal lines represent the spatial images of these fronts after registration in the HiSCORE array. The dashed arrows directed from the top to bottom show the direction of arrival of the narrow pair. It is clear from the left panel, that even if both subevents were truly simultaneous in the sense that their light fronts intersect any plane orthogonal to the direction of motion simultaneously, they would be detected not simultaneously, but with a time shift. The right panel in Fig.~\ref{FigDeltaTScheme} illustrates the opposite situation: subevents can arrive exactly simultaneously in the HiSCORE plane but not be truly simultaneous. It is the true time interval between the events that is of interest, since it is this time that will represent the difference in the emission times of gamma quanta from a point source. Consequently, the geometry of narrow pairs must be recalculated to obtain the true difference in the arrival times of the events.

This recalculation has been carried out. The corresponding true time intervals between components are given in the $\Delta t_{\mathrm{Corrected}}$ column of the Tab.~\ref{TabEvents}. The recalculation procedure not only yields the true time intervals between subevents, but also produces the statistical errors in determining these intervals. The corresponding errors are given in the $\sigma(\Delta t_{\mathrm{Corrected}})$ column of the Tab.~\ref{TabEvents}. It is evident that these accuracies are no worse than 2\,ns, and this is exactly the accuracy with which the simultaneity of subevents in an experiment can be established. But, as can be seen from Tab.~\ref{TabEvents}, the actual time intervals between the components are not units of nanoseconds, but thousands of nanoseconds.

The correspondence between the detected number of narrow paired events in interval $0 < \Delta t < 6\,\mu\mathrm{s}$ to background expectations shows that there is no indication of the existence of an anomalous flow of narrow groups of events for small time windows.

\section{Discussion and conclusion}
\label{CONCLUSION}

In Section~\ref{GG}, we showed that the number of detected narrow spatiotemporal pairs of events corresponds to the expected background of the chance EAS coincidences. This essentially indicates a correct understanding of the background formation for narrow spatiotemporal groups of events. However, the most reliable signature of real primordial black hole evaporation is not the dectection of narrow spatiotemporal pairs, but the detection of narrow groups of events containing 3, 4, or more single events, since the random EAS background for such groups is very small. We present here estimates of this background and corresponding estimates of the statistical reliability of detecting black hole evaporation remnants.

It is difficult to directly estimate the probability of the chance generation of triple, quadruple and more narrow spatiotemporal groups by the EAS background using the Monte Carlo method, as we did in Section \ref{GG} for narrow pairs, since the corresponding probabilities are very small and would require an excessively large number of Monte Carlo tests. We will estimate these probabilities approximately, with an accuracy up to a factor of unity. For example, to estimate the probability of the random appearance of a narrow triple group, we will assume that a narrow double pair has already appeared by chance and, in addition, within the same time interval $\Delta t$ used, third event has appeared that points to the same point in the sky as the first two ones. The probability of the chance appearance of an event in the interval $\Delta t$ is $p_t=f\Delta t$, where $f$ is the average event counting rate. For the 2022--2023 season $f=32.3\,\mathrm{s}^{-1}$, for the 2023--2024 season $f=37.1\,\mathrm{s}^{-1}$ (for events with size not less 5 optical stations). The probability of a chance coincidence of the direction of two events $p_s$ was calculated by the Monte Carlo method using the probability distributions in Fig.~\ref{FigSkyMap}. For the 2022--2023 season $p_s = 3.7\cdot10^{-5}$, for the 2023--2024 season $p_s = 3.4\cdot10^{-5}$. If the background of double narrow groups was $N_2$ for some interval $\Delta t$, then for the probability of the appearance of at least one triple narrow group we get $P_3 \sim N_2\times(p_tp_s)$, for a quadruple narrow group $P_4 \sim N_2\times(p_tp_s)^2$ and so on.

\begin{table}
\centering
\caption{Probabilities of obtaining narrow groups of 3 and 4 events due to chance coincidences of the EAS background for different sizes of time group intervals $\Delta t$ and the maximum distances of detection of the black hole evaporation.\label{TabN3N4}}
\begin{tabular}{|r|c|c|c|c|}
\hline
$\Delta t\,(\mu\mathrm{s})$ & $P_3$ & $P_4$ & $N_\gamma$ & $R$\,(l.y.) \\
\hline
10000 & $4\cdot10^{-2}\,(2.1\sigma)$ & $5\cdot10^{-7}\,(5.0\sigma)$  &   4   &   0.342  \\
 1000 & $4\cdot10^{-4}\,(3.5\sigma)$ & $5\cdot10^{-10}\,(6.2\sigma)$ &   4   &   0.187  \\
  100 & $4\cdot10^{-6}\,(4.6\sigma)$ & $5\cdot10^{-13}\,(7.2\sigma)$ &   4   &   0.091  \\
    6 & $7\cdot10^{-9}\,(5.8\sigma)$ & $-$                           &   3   &   0.036  \\
\hline
\end{tabular}
\end{table}

Probabilities of obtaining narrow groups of 3 and 4 events due to chance coincidences of the EAS background for different sizes of time group intervals $\Delta t$ calculated for the season 2022--2023 are shown in Tab.~\ref{TabN3N4}. The number of narrow background pairs for the interval $\Delta t = 10\,\mathrm{ms}$ was not explicitly determined from the measurement data in this work and was not directly simulated by the Monte Carlo method, but for a rough estimate we can assume that this number will be approximately 10 times greater than for $\Delta t = 1\,\mathrm{ms}$. To calculate the row of Tab.~\ref{TabN3N4} for $\Delta t = 10\,\mathrm{ms}$, the value $N_2 = 3380$ was adopted.

It can be seen from Tab.~\ref{TabN3N4} that the detection of a narrow group with size of 3 events may be simulated by a chance coincidence of EAS only with small probability, therefore it gives a strong indication of a real narrow group of astrophysical origin. The detection of a narrow group with size of 4 events is in all cases a signature of black hole evaporation with a statistical reliability better than five standard deviations. We found no triple of quadruple narrow spatiotemporal groups in the data, therefore we have no indication of black hole evaporation.

Assuming that the black hole radiation spectrum is Planckian one and that it consists solely of gamma rays, it is easy to calculate the total number of gamma rays with energies above 100\,TeV emitted for any time interval before the black hole's complete evaporation. Assuming that the expected number of gamma rays incident on the HiSCORE array is $N_\gamma$ (column $N_\gamma$ in Tab.~\ref{TabN3N4}), it is easy to estimate the maximum distance at which black hole evaporation can be detected. These distances are given in the $R$\,(l.y.) column of Tab.~\ref{TabN3N4}.

It can be seen that the range of detection of the evaporation of the black hole remnant increases with an increase of the time interval $\Delta t$, used to identify narrow spatiotemporal groups of events. In this paper, the task was not to find the optimal interval $\Delta t$ at which the detection range reaches its maximum. This problem will be solved in future works.

Negative results in the search for submicrosecond optical transients and narrow spatiotemporal groups of events allow us to establish new upper limits on the fluxes of the corresponding events. Over six seasons of optical transient searches with the HiSCORE array, starting from the 2018-2019 season to 2023-2024, an exposure of 3478 hours was accumulated, which, assuming a field of view of the instrument of 0.6 steradians (as was already done in \cite{PANOV-2025}), gives an upper limit on the flux of optical transients with a brightness of at least $10^{-4}\mathrm{erg/s/cm}^2$ as $5\cdot10^{-4}\,\mathrm{ster}^{-1}\mathrm{h}^{-1}$. Over two seasons of searching for narrow spatiotemporal groups of gamma quanta (2022--2023, 2023--2024), an exposure of 1356 hours was accumulated, from which we obtain a limit on the flux of narrow spatiotemporal groups of quanta of $10^{-3}\,\mathrm{ster}^{-1}\mathrm{h}^{-1}$.

% \bibliography{HiSCORE2025Short.bib}% common bib file

\begin{thebibliography}{21}
\ifx \bisbn   \undefined \def \bisbn  #1{ISBN #1}\fi
\ifx \binits  \undefined \def \binits#1{#1}\fi
\ifx \bauthor  \undefined \def \bauthor#1{#1}\fi
\ifx \batitle  \undefined \def \batitle#1{#1}\fi
\ifx \bjtitle  \undefined \def \bjtitle#1{#1}\fi
\ifx \bvolume  \undefined \def \bvolume#1{\textbf{#1}}\fi
\ifx \byear  \undefined \def \byear#1{#1}\fi
\ifx \bissue  \undefined \def \bissue#1{#1}\fi
\ifx \bfpage  \undefined \def \bfpage#1{#1}\fi
\ifx \blpage  \undefined \def \blpage #1{#1}\fi
\ifx \burl  \undefined \def \burl#1{\textsf{#1}}\fi
\ifx \doiurl  \undefined \def \doiurl#1{\url{https://doi.org/#1}}\fi
\ifx \betal  \undefined \def \betal{\textit{et al.}}\fi
\ifx \binstitute  \undefined \def \binstitute#1{#1}\fi
\ifx \binstitutionaled  \undefined \def \binstitutionaled#1{#1}\fi
\ifx \bctitle  \undefined \def \bctitle#1{#1}\fi
\ifx \beditor  \undefined \def \beditor#1{#1}\fi
\ifx \bpublisher  \undefined \def \bpublisher#1{#1}\fi
\ifx \bbtitle  \undefined \def \bbtitle#1{#1}\fi
\ifx \bedition  \undefined \def \bedition#1{#1}\fi
\ifx \bseriesno  \undefined \def \bseriesno#1{#1}\fi
\ifx \blocation  \undefined \def \blocation#1{#1}\fi
\ifx \bsertitle  \undefined \def \bsertitle#1{#1}\fi
\ifx \bsnm \undefined \def \bsnm#1{#1}\fi
\ifx \bsuffix \undefined \def \bsuffix#1{#1}\fi
\ifx \bparticle \undefined \def \bparticle#1{#1}\fi
\ifx \barticle \undefined \def \barticle#1{#1}\fi
\bibcommenthead
\ifx \bconfdate \undefined \def \bconfdate #1{#1}\fi
\ifx \botherref \undefined \def \botherref #1{#1}\fi
\ifx \url \undefined \def \url#1{\textsf{#1}}\fi
\ifx \bchapter \undefined \def \bchapter#1{#1}\fi
\ifx \bbook \undefined \def \bbook#1{#1}\fi
\ifx \bcomment \undefined \def \bcomment#1{#1}\fi
\ifx \oauthor \undefined \def \oauthor#1{#1}\fi
\ifx \citeauthoryear \undefined \def \citeauthoryear#1{#1}\fi
\ifx \endbibitem  \undefined \def \endbibitem {}\fi
\ifx \bconflocation  \undefined \def \bconflocation#1{#1}\fi
\ifx \arxivurl  \undefined \def \arxivurl#1{\textsf{#1}}\fi
\csname PreBibitemsHook\endcsname

%%% 1
\bibitem[\protect\citeauthoryear{Abdo et~al.}{2015}]{Milagro-2015}
\begin{barticle}
\bauthor{\bsnm{Abdo}, \binits{A.A.}},
\bauthor{\bsnm{Abeysekara}, \binits{A.U.}},
\bauthor{\bsnm{Alfaro}, \binits{R.}},
\bauthor{\bsnm{Allen}, \binits{B.T.}},
\bauthor{\bsnm{Alvarez}, \binits{C.}}, \betal:
\batitle{Milagro limits and {HAWC} sensitivity for the rate-density of
  evaporating primordial black holes}.
\bjtitle{Astropart. Phys.}
\bvolume{64},
\bfpage{4}--\blpage{12}
(\byear{2015})
\doiurl{10.1016/j.astropartphys.2014.10.007}
{\href{https://arxiv.org/abs/arXiv:1407.1686}{{arXiv:1407.1686}}}
\end{barticle}
\endbibitem

%%% 2
\bibitem[\protect\citeauthoryear{Albert et~al.}{2018}]{HAWC-2020}
\begin{barticle}
\bauthor{\bsnm{Albert}, \binits{A.}},
\bauthor{\bsnm{Alfaro}, \binits{R.}},
\bauthor{\bsnm{Alvarez}, \binits{C.}},
\bauthor{\bsnm{Arteaga-Vel{\'a}zquez}, \binits{J.C.}},
\bauthor{\bsnm{Arunbabu}, \binits{K.P.}}, \betal:
\batitle{Constraining the local burst rate density of primordial black holes
  with hawc}.
\bjtitle{ApJ}
\bvolume{857},
\bfpage{49}
(\byear{2018})
\doiurl{10.1088/1475-7516/2020/04/026}
{\href{https://arxiv.org/abs/arXiv:1911.04356}{{arXiv:1911.04356}}}
\end{barticle}
\endbibitem

%%% 3
\bibitem[\protect\citeauthoryear{Aharonian et~al.}{2023}]{HESS-2023}
\begin{barticle}
\bauthor{\bsnm{Aharonian}, \binits{F.}},
\bauthor{\bsnm{{Ait Benkhali}}, \binits{F.}},
\bauthor{\bsnm{Aschersleben}, \binits{J.}},
\bauthor{\bsnm{B{\"o}ttcher}, \binits{M.}},
\bauthor{\bsnm{Backes}, \binits{M.}},
\bauthor{\bsnm{{et al. (H. E. S. S. Collaboration)}}}:
\batitle{Search for the evaporation of primordial black holes with h.e.s.s.}
\bjtitle{JCAP}
\bvolume{04},
\bfpage{040}
(\byear{2023})
\doiurl{10.1088/1475-7516/2023/04/040}
{\href{https://arxiv.org/abs/arXiv:2303.12855}{{arXiv:2303.12855}}}
\end{barticle}
\endbibitem

%%% 4
\bibitem[\protect\citeauthoryear{Alexandreas et~al.}{1993}]{CYGNUS-1993}
\begin{barticle}
\bauthor{\bsnm{Alexandreas}, \binits{D.E.}},
\bauthor{\bsnm{Allen}, \binits{G.E.}},
\bauthor{\bsnm{Berley}, \binits{D.}},
\bauthor{\bsnm{Biller}, \binits{S.}},
\bauthor{\bsnm{Burman}, \binits{R.L.}}, \betal:
\batitle{New limit on the rate-density of evaporating black holes}.
\bjtitle{Phys. Rev. Lett.}
\bvolume{71},
\bfpage{2524}--\blpage{2527}
(\byear{1993})
\doiurl{10.1103/PhysRevLett.71.2524}
\end{barticle}
\endbibitem

%%% 5
\bibitem[\protect\citeauthoryear{Ackermann et~al.}{2018}]{FERMI-2018}
\begin{barticle}
\bauthor{\bsnm{Ackermann}, \binits{M.}},
\bauthor{\bsnm{Atwood}, \binits{W.B.}},
\bauthor{\bsnm{Baldini}, \binits{L.}},
\bauthor{\bsnm{Ballet}, \binits{J.}},
\bauthor{\bsnm{Barbiellini}, \binits{G.}}, \betal:
\batitle{Search for gamma-ray emission from local primordial black holes with
  the fermi large area telescope}.
\bjtitle{ApJ}
\bvolume{857},
\bfpage{49}
(\byear{2018})
\doiurl{10.3847/1538-4357/aaac7b}
{\href{https://arxiv.org/abs/arXiv:1802.00100}{{arXiv:1802.00100}}}
\end{barticle}
\endbibitem

%%% 6
\bibitem[\protect\citeauthoryear{Amenomori et~al.}{1995}]{Tibet-1995}
\begin{bchapter}
\bauthor{\bsnm{Amenomori}, \binits{M.}},
\bauthor{\bsnm{Cao}, \binits{Z.}},
\bauthor{\bsnm{Dai}, \binits{B.Z.}},
\bauthor{\bsnm{Ding}, \binits{L.K.}},
\bauthor{\bsnm{Feng}, \binits{Y.X.}}, \betal:
\bctitle{Search for 10 {TeV} gamma bursts from evaporating primordial black
  holes with the {T}ibet air shower array}.
In: \bbtitle{24th International Cosmic Ray Conference},
vol. \bseriesno{2},
p. \bfpage{112}
(\byear{1995})
\end{bchapter}
\endbibitem

%%% 7
\bibitem[\protect\citeauthoryear{{Archambault, S. et al. (VERITAS
  Collaboration)}}{2017}]{VERITAS-2017}
\begin{barticle}
\bauthor{\bsnm{{Archambault, S. et al. (VERITAS Collaboration)}}}:
\batitle{Search for primordial black hole evaporation with veritas}.
\bjtitle{Proceedings of Science}
\bvolume{ICRC2017},
\bfpage{691}
(\byear{2017})
{\href{https://arxiv.org/abs/arXiv:1709.00307}{{arXiv:1709.00307}}}
\end{barticle}
\endbibitem

%%% 8
\bibitem[\protect\citeauthoryear{Budnev et~al.}{2017}]{TAIGA-NIM-2017A}
\begin{barticle}
\bauthor{\bsnm{Budnev}, \binits{N.}},
\bauthor{\bsnm{Astapov}, \binits{I.}},
\bauthor{\bsnm{Barbashina}, \binits{N.}},
\bauthor{\bsnm{Barnyakov}, \binits{A.}},
\bauthor{\bsnm{Bezyazeekov}, \binits{P.}}, \betal:
\batitle{The {TAIGA} experiment: {F}rom cosmic-ray to gamma-ray astronomy in
  the {T}unka valley}.
\bjtitle{Nuclear Inst. and Methods in Physics Research, A}
\bvolume{845},
\bfpage{330}--\blpage{333}
(\byear{2017})
\doiurl{10.1016/j.nima.2016.06.041}
\end{barticle}
\endbibitem

%%% 9
\bibitem[\protect\citeauthoryear{Budnev et~al.}{2020}]{TAIGA-NIM-2020B}
\begin{barticle}
\bauthor{\bsnm{Budnev}, \binits{N.}},
\bauthor{\bsnm{Astapov}, \binits{I.I.}},
\bauthor{\bsnm{Bezyazeekov}, \binits{P.A.}},
\bauthor{\bsnm{Borodin}, \binits{A.}},
\bauthor{\bsnm{Br{\"u}ckner}, \binits{M.}}, \betal:
\batitle{{TAIGA} -- {A} hybrid array for high-energy gamma astronomy and
  cosmic-ray physics}.
\bjtitle{Nuclear Inst. and Methods in Physics Research, A}
\bvolume{958},
\bfpage{162113}
(\byear{2020})
\doiurl{10.1016/j.nima.2019.04.067}
\end{barticle}
\endbibitem

%%% 10
\bibitem[\protect\citeauthoryear{Blandford}{1977}]{BLANDFORD-1977}
\begin{barticle}
\bauthor{\bsnm{Blandford}, \binits{R.D.}}:
\batitle{Spectrum of a radio pulse from an exploding black hole.}
\bjtitle{MNRAS}
\bvolume{181},
\bfpage{489}--\blpage{498}
(\byear{1977})
\doiurl{10.1093/mnras/181.3.489}
\end{barticle}
\endbibitem

%%% 11
\bibitem[\protect\citeauthoryear{Bugaev et~al.}{2009a}]{PETKOV-2009B}
\begin{botherref}
\oauthor{\bsnm{Bugaev}, \binits{E.V.}},
\oauthor{\bsnm{Petkov}, \binits{V.B.}},
\oauthor{\bsnm{Gaponenko}, \binits{A.N.}},
\oauthor{\bsnm{Klimai}, \binits{P.A.}},
\oauthor{\bsnm{Andreev}, \binits{M.V.}}, et al.:
Experimental search of bursts of gamma rays from primordial black holes using
  different evaporation models
(2009)
\end{botherref}
\endbibitem

%%% 12
\bibitem[\protect\citeauthoryear{Bugaev et~al.}{2009b}]{PETKOV-2009A}
\begin{botherref}
\oauthor{\bsnm{Bugaev}, \binits{E.V.}},
\oauthor{\bsnm{Petkov}, \binits{V.B.}},
\oauthor{\bsnm{Gaponenko}, \binits{A.N.}},
\oauthor{\bsnm{Klimai}, \binits{P.A.}},
\oauthor{\bsnm{Andreev}, \binits{M.V.}}, et al.:
Experimental search of bursts of very high energy gamma rays from primordial
  black holes
(2009)
\end{botherref}
\endbibitem

%%% 13
\bibitem[\protect\citeauthoryear{Gress et~al.}{2017}]{TAIGA-NIM-2017B}
\begin{barticle}
\bauthor{\bsnm{Gress}, \binits{O.}},
\bauthor{\bsnm{Astapov}, \binits{I.}},
\bauthor{\bsnm{Budnev}, \binits{N.}},
\bauthor{\bsnm{Bezyazeekov}, \binits{P.}},
\bauthor{\bsnm{Bogdanov}, \binits{A.}}, \betal:
\batitle{The wide-aperture gamma-ray telescope {TAIGA}-{HiSCORE} in the {T}unka
  {V}alley: {D}esign, composition and commissioning}.
\bjtitle{Nuclear Inst. and Methods in Physics Research, A}
\bvolume{845},
\bfpage{367}--\blpage{372}
(\byear{2017})
\doiurl{10.1016/j.nima.2016.08.031}
\end{barticle}
\endbibitem

%%% 14
\bibitem[\protect\citeauthoryear{Kuzmichev et~al.}{2020}]{TAIGA-NIM-2020A}
\begin{barticle}
\bauthor{\bsnm{Kuzmichev}, \binits{L.}},
\bauthor{\bsnm{Astapov}, \binits{I.}},
\bauthor{\bsnm{Bezyazeekov}, \binits{P.}},
\bauthor{\bsnm{Borodin}, \binits{A.}},
\bauthor{\bsnm{Br{\"u}ckner}, \binits{M.}}, \betal:
\batitle{Cherenkov {EAS} arrays in the {T}unka astrophysical center: {F}rom
  {T}unka-133 to the {TAIGA} gamma and cosmic ray hybrid detector}.
\bjtitle{Nuclear Inst. and Methods in Physics Research, A}
\bvolume{952},
\bfpage{161830}
(\byear{2020})
\doiurl{10.1016/j.nima.2019.01.056}
\end{barticle}
\endbibitem

%%% 15
\bibitem[\protect\citeauthoryear{Linton et~al.}{2006}]{Whipple-2006}
\begin{barticle}
\bauthor{\bsnm{Linton}, \binits{E.T.}},
\bauthor{\bsnm{Atkins}, \binits{R.W.}},
\bauthor{\bsnm{Badran}, \binits{H.M.}},
\bauthor{\bsnm{Blaylock}, \binits{G.}},
\bauthor{\bsnm{Boyle}, \binits{P.J.}}, \betal:
\batitle{A new search for primordial black hole evaporations using the whipple
  gamma-ray telescope}.
\bjtitle{JCAPS}
\bvolume{01},
\bfpage{013}
(\byear{2006})
\doiurl{10.1088/1475-7516/2006/01/013}
\end{barticle}
\endbibitem

%%% 16
\bibitem[\protect\citeauthoryear{Panov et~al.}{2021}]{PANOV-2021}
\begin{barticle}
\bauthor{\bsnm{Panov}, \binits{A.D.}},
\bauthor{\bsnm{Astapov}, \binits{I.I.}},
\bauthor{\bsnm{Beskin}, \binits{G.M.}},
\bauthor{\bsnm{Bezyazeekov}, \binits{P.A.}},
\bauthor{\bsnm{Blank}, \binits{M.}}, \betal:
\batitle{Search for astrophysical nanosecond optical transients with
  taiga-hiscore array}.
\bjtitle{Physics of Atomic Nuclei}
\bvolume{84},
\bfpage{1037}--\blpage{1044}
(\byear{2021})
\doiurl{10.1134/S1063778821130251}
{\href{https://arxiv.org/abs/arXiv:2109.09637}{{arXiv:2109.09637}}}
\end{barticle}
\endbibitem

%%% 17
\bibitem[\protect\citeauthoryear{Panov et~al.}{2025}]{PANOV-2025}
\begin{barticle}
\bauthor{\bsnm{Panov}, \binits{A.D.}},
\bauthor{\bsnm{Astapov}, \binits{I.I.}},
\bauthor{\bsnm{Beskin}, \binits{G.M.}},
\bauthor{\bsnm{Bezyazykov}, \binits{P.A.}},
\bauthor{\bsnm{Blinov}, \binits{A.V.e.}}:
\batitle{Four years of wide-field search for nanosecond optical transients with
  the {TAIGA-HiSCORE} {C}herenkov array}.
\bjtitle{Astronomy Reports}
\bvolume{69},
\bfpage{433}--\blpage{456}
(\byear{2025})
\doiurl{10.1134/S106377292570177X}
{\href{https://arxiv.org/abs/arXiv:2412.00159}{{arXiv:2412.00159}}}
\end{barticle}
\endbibitem

%%% 18
\bibitem[\protect\citeauthoryear{Petkov et~al.}{2008}]{PETKOV-2008}
\begin{barticle}
\bauthor{\bsnm{Petkov}, \binits{V.B.}},
\bauthor{\bsnm{Bugaev}, \binits{E.V.}},
\bauthor{\bsnm{Klimai}, \binits{P.A.}},
\bauthor{\bsnm{Andreev}, \binits{M.V.}},
\bauthor{\bsnm{Volchenko}, \binits{V.I.}}, \betal:
\batitle{Searching for very-high-energy gamma-ray bursts from evaporating
  primordial black holes}.
\bjtitle{Astronomy Letters}
\bvolume{34},
\bfpage{509}--\blpage{514}
(\byear{2008})
\doiurl{10.1134/S106377370808001X}
{\href{https://arxiv.org/abs/arXiv.0808.3093}{{arXiv.0808.3093}}}
\end{barticle}
\endbibitem

%%% 19
\bibitem[\protect\citeauthoryear{Petkov et~al.}{2019}]{PETKOV-2019}
\begin{botherref}
\oauthor{\bsnm{Petkov}, \binits{V.B.}},
\oauthor{\bsnm{Bugaev}, \binits{E.V.}},
\oauthor{\bsnm{Klimai.}, \binits{P.A.}}:
Multimessenger search for evaporating primordial black holes.
(2019)
\end{botherref}
\endbibitem

%%% 20
\bibitem[\protect\citeauthoryear{Rees}{1977}]{REES-1977}
\begin{barticle}
\bauthor{\bsnm{Rees}, \binits{M.J.}}:
\batitle{A better way of searching for black-hole explosions?}
\bjtitle{Nature}
\bvolume{266},
\bfpage{333}--\blpage{334}
(\byear{1977})
\doiurl{10.1038/266333a0}
\end{barticle}
\endbibitem

%%% 21
\bibitem[\protect\citeauthoryear{Tluczykont et~al.}{2011}]{HISCORE-2011}
\begin{barticle}
\bauthor{\bsnm{Tluczykont}, \binits{M.}},
\bauthor{\bsnm{Hampf}, \binits{D.}},
\bauthor{\bsnm{Horns}, \binits{D.}},
\bauthor{\bsnm{Kneiske}, \binits{T.}},
\bauthor{\bsnm{Eichler}, \binits{R.}},
\bauthor{\bsnm{Nachtigall}, \binits{R.}},
\bauthor{\bsnm{Rowell}, \binits{G.}}:
\batitle{The ground-based large-area wide-angle gamma-ray and cosmic-ray
  experiment {HiSCORE}}.
\bjtitle{Adv. Space Res.}
\bvolume{48},
\bfpage{1935}--\blpage{1941}
(\byear{2011})
\doiurl{10.1016/j.asr.2011.08.004}
\end{barticle}
\endbibitem

\end{thebibliography}

\end{document}